\documentclass[reprint,amsmath,amssymb,aps,prl]{revtex4-2}

\usepackage{graphicx}
\usepackage{dcolumn}
\usepackage{bm}
\usepackage{braket}
\usepackage{xcolor}
\usepackage{bbold}

\begin{document}

 \title{Identifying differences between semi-classical and full-quantum descriptions of plexcitons}
\author{Marco Romanelli}
\affiliation{Department of Chemical Sciences, University of Padova, via Marzolo 1, 35131 Padova, Italy}
\author{Stefano Corni}
\email{stefano.corni@unipd.it}
\affiliation{Department of Chemical Sciences, University of Padova, via Marzolo 1, 35131 Padova, Italy}
\affiliation{CNR Institute of Nanoscience, via Campi 213/A, 41125 Modena, Italy}
\affiliation{Padua Quantum Technologies Research Center, University of Padova, 35131 Padova, Italy}

\begin{abstract}
 Strong light-matter coupling between molecules and plasmonic nanoparticles give rise to new hybrid eigenstates of the coupled system, commonly referred to as polaritons, or more precisely, plexcitons. Over the last decade it has been amply shown that molecular electron dynamics and photophysics can be drastically affected by such interactions, thus paving the way for light-induced control of molecular excited-state properties and reactivity. Here, by combining \textit{ab initio} molecular description and classical or quantum modelling of arbitrarily-shaped plasmonic nanostructures within Stochastic Schr\"{o}dinger Equation, we present two approaches, one semi-classical and one full-quantum, to follow in real-time the electronic dynamics of plexcitons while realistically taking plasmonic dissipative losses into account. The full-quantum theory is compared with the semi-classical analogue under different interaction regimes, showing (numerically and theoretically) that even in the weak-field and weak-coupling limit a small-yet-observable difference arises. 
 
\end{abstract}

\maketitle
 
\section{Introduction}
Plexcitonic systems, namely nanohybrid architectures composed of plasmonic nanostructures interacting with molecular species, have been drawing ever-increasing attention over the past few years since they proved to be a non-invasive way of changing molecular properties as a result of light-matter coupling\cite{zhao2022,ma2021,manuel2019,thomas2018,fregoni2022,feist2018}. Indeed, many recent works have illustrated the possibility of using plasmonic platforms to affect not only absorption and emission properties of photo-active molecules\cite{lee2019,benz2016,imada2017single,yang2020,lakowicz2004,lakowicz2005,anger2006enhancement,della2013handbook,romanelli2021}, but also energy transfer rates\cite{kong2022,cao2021,georgiou2018control,du2018theory,coles2014}, photorelaxation channels\cite{antoniou2020,munkhbat2018,felicetti2020,gu2020,torres2021molecular,kuttruff2022} and photochemical reactions\cite{fregoni2022,fregoni2020,fregoni2018,riso2022,li2018,gao2011,linic2015,torres2021}, just to mention a few. \\
\indent The degree of coupling between molecular emitters and plasmonic resonators in such cases can span different regimes, being defined "weak" when it is small with respect to the dissipative losses of the coupled system, or "strong" in the opposite scenario\cite{vasa2018,tame2013}. Usually,  in the former case  perturbative semiclassical approaches are believed to suffice to capture the modified molecular response due to the weak plasmon-molecule interaction and they have been widely used to account for enhancement (Purcell effect) or suppression of radiative molecular emission, quenching of molecular excited states lifetimes and also molecular excitation energies shifts  because of the nearby plasmonic nanoparticles (NPs)\cite{vukovic2009,caricato2006,corni2003,mennucci2019,hohenester2008,lopata2009,chen2010,zhang2020,aguilar2019} (known as "medium-induced Lamb shift"). \\
On the other hand, when the light-matter coupling is large enough to exceed the dissipative losses of the coupled system, and the molecular and plasmonic excitations are resonant, new hybrid molecular-plasmonic eigenstates, commonly named plexcitons, are actually formed, resulting in a coherent energy exchange between the molecular excited state and the plasmonic system, thus defining the onset of the "strong coupling" regime\cite{fregoni2018,reshmi2018,vasa2018,peruffo2021}. \\
\indent Since the seminal work of Hutchison et al.\cite{hutchison2012}, controlling photochemical reaction rates and photocatalytic processes thanks to confined light modes has given rise to a vibrant and active area of research. If, on the one hand, plasmonic nanocavities enable the confinement of light in sub-nanometric volumes, thus boosting the light-matter interaction to such an extent that even single-molecule strong coupling becomes feasible\cite{chikkaraddy2016single}, on the other they are typically associated to large (fast) dissipative losses because of the well-known ultrafast plasmon dephasing process, happening on a femtosecond (fs) time scale, which quickly leads to a non-radiative dissipation of the initial plasmonic excitation\cite{stockman2007,hartland2011,koya2022}. Since these processes are commonly faster than usual electronic processes taking place in photo-excited molecules, tailoring them for chemical applications calls for theoretical models able to describe those dynamical  interactions, whether they are "weak" or "strong", while realistically taking account of such dissipative losses, as they can drastically affect the resulting molecular electron dynamics and thus being impactful for possible applications. \\
\indent In the following, building on the previously-developed modelling strategy aimed at describing in real-time the electron dynamics of molecules close to classically-described plasmonic nanostructures\cite{coccia2019,coccia2020,pipolo2016}, hereafter labelled as  semi-classical (SC) approach, we push that theory one step further to plexcitonic wavefunctions, based on the quantized description of the plasmonic response\cite{fregoni2021}. The latter picture is referred to in the following as full-quantum model (FQ).\\ 
 \indent Limiting ourselves to time-dependent modelling, different approaches based on semiclassical Maxwell-Bloch equations, coupled harmonic oscillator models, density-matrix propagation through master equations and Heisenberg-Langevin equations have been used before\cite{li2019,waks2010,pelton2019,yang2019,cuartero2018,cortes2020,nascimento2015,davidsson2020,chen2013,kaminski2007,fofang2011,zhang2006,hoffmann2019,rokaj2018,schafer2022,chen2023}, but in the majority of those cases the molecules are simply described as two-state quantum emitters and the metallic NPs that are considered are typically characterized by simple shapes for which analytical solutions of the scattering Green's function are easily available. In this context, full-quantum models rooted in macroscopic quantum electrodynamics (QED) have also been used to investigate the population dynamics of multiple two-state emitters coupled to complex bath spectral densities representing realistic metal environments \cite{feist2020,medina2021,sanchez2022,trugler2008}. Other macroscopic QED-based approaches investigated molecular emission features in the presence of planar metallic mirrors\cite{wang2019,wang2020jcp,wang2020jcpl,chuang2022}, but in all these cases a simplified description of the emitters is considered. On the other hand, real-time \textit{ab initio} investigations based on using Real-Time Time-Dependent Density Functional Theory (RT-TDDFT) for the full system have recently started to emerge, but due to the high computational cost of such simulations only small metal clusters consisting of at most tens of atoms have been described, thus limiting a direct comparison with realistic experimental  setups\cite{sakko2014,rossi2019,kuisma2022,fojt2021,castro2004optical,martinez2006photoabsorption,falke2014coherent,lucchini2021unravelling,ruggenthaler2018quantum,varas2016quantum,zhang2018,wu2023}.In particular, none of these models couple an atomistic quantum description of real molecular structures with NPs of arbitrary shape and dimension, thus tackling systems of real complexity and practical usage. This is the core feature of the methods that we hereby present, which combine state of the art quantum chemistry description of molecules with classical (SC) or quantum modelling (FQ) of arbitrarily-shaped plasmonic NPs, laying the groundwork for  a direct comparison of the two regimes on an equal footing. \\
Dissipative losses of the system are treated using a Stochastic Schr\"{o}dinger Equation (SSE) formalism, which is an alternative approach to density-matrix based propagation that focuses on following directly in time the system wave-function evolution under the influence of the surroundings\cite{biele2012,coccia2018}. \\
\indent Previous works\cite{torma2014,waks2010,slowik2013} have shown that the predicted total absorbed power by dipolar emitters coupled to spherical plasmonic NPs differ between semi-classical and full-quantum descriptions under high intensity driving fields. More precisely, results derived from semi-classical Maxwell-Bloch equations leads to an overestimation of the system absorbed power compared to the exact full-quantum results obtained by full master equation propagation. The origin of this divergence has been related to non-linear effects that arise upon exciting the system with high-intensity fields because of emitters saturation and optical bistability\cite{savage1988,szoke1969,rice1994,drummond1980}. On the other hand, when the weak-coupling and weak-field limits\cite{waks2006,waks2010,vogell2013} do apply, the two approaches are expected to give the same results, even in the presence of environment-induced dissipation\cite{waks2006,waks2010}. \\
In the following, by providing a direct comparison between the two descriptions in a system composed of a plasmonic NP and a molecule, we do numerically confirm the expected divergence under strong-field excitation, but we also observe a slight difference in the molecular excited state population upon external driving when the molecular and plasmon systems are not resonant, even under linear excitation regimes. We find out that the origin of this discrepancy is intimately connected to the anti-resonant term of the NP linear response polarizability, which enters in different ways in the SC and FQ models. \\ 
It is worth pointing out that these two approaches are theoretically and numerically comparable since they share the same theoretical ingredients (the numerical response to an external oscillating and spatially varying electric field is identical for both by construction), thus allowing to pinpoint fundamental differences in the way plasmon-molecules interactions are described in the two cases.

\section{Results and discussion}\label{Results}
In the SC picture, the molecule is described at quantum mechanical level, but the plasmonic NP is treated as a classical polarizable continuum object with the PCM-NP model, which has been previously developed in our group\cite{mennucci2019,pipolo2016}. It essentially relies on solving the electromagnetic problem of coupling a quantum chemistry molecular description with the nearby homogeneous plasmonic system by numerically solving the corresponding Poisson's equation through a Boundary-Element-Method\,(BEM)\cite{de2002} approach. The NP response to external perturbations (e.g. molecular electron densities or external fields) is expressed in terms of surface charges liying on the NP discretized surface whose discretization is needed to numerically solve the BEM problem (more details can be found in SI 1.1). In this picture, the system Hamiltonian $\hat{H}_{S}(t)$, here renamed $\hat{H}_{SC}(t)$ for the actual SC case, reads:
 \begin{equation}\label{eq:Hsc}
\hat{H}_{SC}(t)=\hat{H}_{mol}-\vec{\hat{\mu}}\cdot\vec{E}_{ext}(t)+(\textbf{q}_{ref}(t)+\textbf{q}_{pol}(t))\cdot\hat{\textbf{V}}
\end{equation}

 \noindent where $\hat{H}_{mol}$ is the time-independent molecular Hamiltonian, $\vec{E}_{ext}(t)$ is the time-dependent external electric field that is used to drive the system, $\vec{\hat{\mu}}$ is the molecular dipole operator, $\textbf{q}_{ref}(t)$ and $\textbf{q}_{pol}(t)$ are vectors collecting the NP response charges on the NP's discretized surface induced by direct polarization of the incoming exciting field ($\textbf{q}_{ref}(t)$) and by the time-dependent nearby molecular electron density ($\textbf{q}_{pol}(t)$), and $\hat{\textbf{V}}$ is the molecular electrostatic potential operator evaluated at the nanoparticle surface where response charges lie on\cite{pipolo2016,mennucci2019}. We point out that the $\textbf{q}_{pol}(t)$ term leads to a non-linear self-interaction effect because those response charges are induced on the NP by the presence of the nearby molecular density and can in turn generate an electric field that can act back on the molecule itself. In the SC limit, under the quasi-static approximation, the imaginary component of this self-interaction contribution leads to an additional non-radiative decay rate for molecular excited states, representing energy transfer to the NP, where the excitation is then quickly dissipated\cite{corni2003,hohenester2008}. This decay process is typically faster than any other intrinsic molecular decay rate\cite{andreussi2004,romanelli2021} when molecules are very close to metallic NPs ($< 1\,\text{nm}$) and so its effect cannot be neglected. Herein, since the main goal of the present work is to compare the SC and FQ models on a perfectly consistent ground, we solely focus on this NP-induced decay channel. Therefore, in the SC picture the wavefunction time-propagation is directly performed with the Hamiltonian $\hat{H}_{SC}(t)$ of eq.\ref{eq:Hsc} and no additional decay operators have to be included (see SI 1.1).

\indent On the other hand, in the FQ picture the plasmonic NP is also quantized, so the system Hamiltonian becomes
\begin{equation}\label{eq:Hfq}
    \hat{H}_{FQ}(t)=\hat{H}_{0,FQ}-\vec{\hat{\mu}}\cdot\vec{E}_{ext}(t)
\end{equation}

\noindent where $\hat{H}_{0,FQ}$  is the full plasmon-molecule Hamiltonian\cite{fregoni2021},

\begin{equation}\label{eq:Hfq_full}
    \hat{H}_{0,FQ}= \hat{H}_{mol}+\sum_p\omega_{p}{\hat{b}}^{\dagger}_{p}{\hat{b}}_{p}+\sum_{pj}q_{pj}\hat{V}_{j}({b}^{\dagger}_{p}+{b}_{p})
\end{equation}


\noindent where $\omega_p$ is the frequency of the $p^{th}$ quantized plasmon mode of the NP and $\hat{b}^{\dagger}_{p},\,\hat{b}_{p}$ are the corresponding plasmonic creation and annihilation operators, respectively. In eq.\ref{eq:Hfq_full}, the \textit{j} index labels the $j^{th}$ surface element  of the NP (called "tessera") after numerical discretization that is needed to solve the BEM equations, leading then to the corresponding quantized surface charge $q_{pj}$ for a given $p^{th}$ plasmon mode. $\hat{V}_{j}$ is instead the molecular electrostatic potential operator evaluated at the $j^{th}$ tessera. \\ 
The full derivation of the Q-PCM-NP quantization scheme has been detailed elsewhere\cite{fregoni2021}. Here it is important to remark that  the quantum model is derived to provide the very same linear response polarization in the nanoparticle as the classical one. Further details on the FQ model can be found in SI 1.2. \\
The wavefunction propagation is then performed in a SSE framework (SI 1.3) to consistently include plasmon-induced losses as in the SC picture. This is achieved by using the following Hamiltonian,  
 \begin{equation}\label{eq:Hsse_fq}
\hat{H}_{SSE,FQ}(t)=\hat{H}_{FQ}(t)-\frac{i}{2}\sum_p\hat{S}_{p,FQ}^{\dagger}\hat{S}_{p,FQ}
\end{equation}

\noindent with
 \begin{equation}\label{eq:S_fq}
\hat{S}_{p,FQ}=\sqrt{\Gamma_p}\mathbb{1_{mol}}\otimes\left(\ket{0}\bra{1_p}\right)
\end{equation}
\noindent where $\mathbb{1_{mol}}$ is the identity operator on the molecular states, and $\Gamma_p$ is the decay rate of the $p^{th}$ quantized mode. 

\begin{figure}[ht!]
    \centering
    \includegraphics[width=8.6cm]{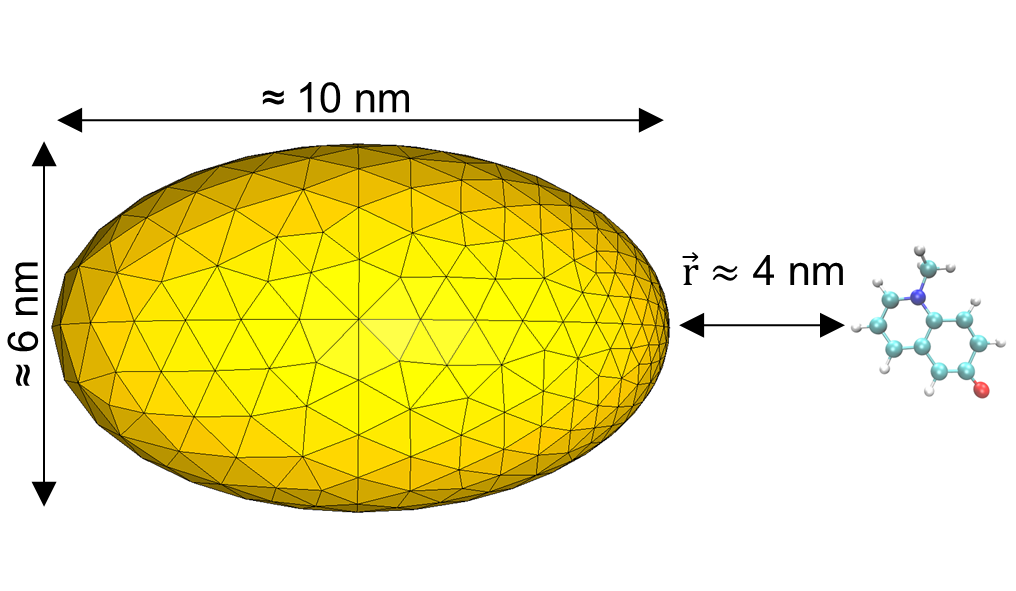}
    \caption{System under investigation composed of a plasmonic gold NP of ellipsoidal shape and N-methyl-6-quinolone molecule. The molecule -NP dimensions are not to scale. }
    \label{fig:setup}
\end{figure}

\indent The SC and FQ models have been compared on a system composed of a plasmonic ellipsoidal NP and N-methyl-6-quinolone molecule (for simplicity we will refer to the latter simply as quinolone), as shown in Fig.\ref{fig:setup}. The choice of such molecular species has been made due to its interesting and previously-investigated excited state properties\cite{klamroth2006}, but for the purpose of the present work other molecules may have been chosen. The quinolone molecule is described at the level of Configuration Interaction Singles (CIS) and only its lowest excited state ($\ket{e}$) is considered in the following for simplicity. The coupled system is driven by a pulse of gaussian shape resonant with the lowest NP plasmon mode frequency $\omega_p = 2.95\,\text{eV}$ (further computational details can be found in SI 2). 

\begin{figure*}[ht!]
    \centering
    \includegraphics[width=1.0\textwidth]{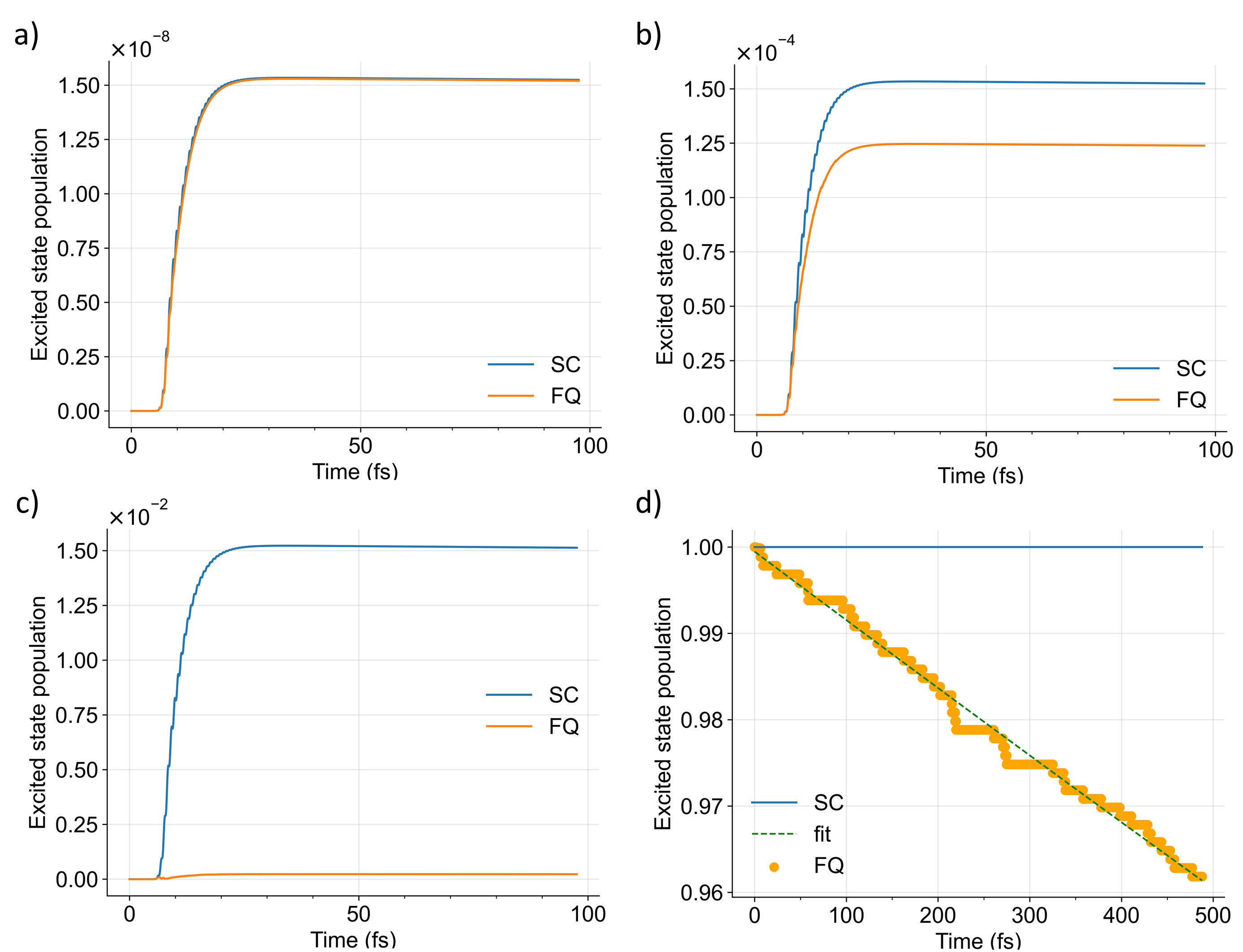}
    \caption{Molecular excited state population over time obtained via SC (blue) and FQ (orange) models under resonance condition ($\delta= \omega_e - \omega_p=0$) for different driving field intensities a)-c). The same setup of Fig.\,\ref{fig:setup} is excited with a Gaussian pulse (SI 2) resonant with the lowest NP plasmon mode $\omega_p=2.95\,\text{eV} $ featuring an intensity of a) $3.5\times10^{4}\,\text{W/cm}^2$, b) $3.5\times10^{8}\,\text{W/cm}^2$ and c) $3.5\times10^{10}\,\text{W/cm}^2$. The extreme limit where the entire molecular population would be in the excited state is reported in d), where the system is initiated in the molecular excited state already at time zero. In this case time-propagation begins from this extreme condition and no driving field is applied. The green dashed line is the result of fitting the corresponding data points with an exponential decay function $f(t)=k e^{-t/\tau}$.}
    \label{fig:high}
\end{figure*}

\indent Under high intensity driving fields the two approaches are expected to diverge\cite{waks2006,szoke1969,rice1994,drummond1980}. An intuitive qualitative explanation of the origin of this divergence can be grasped by considering an oscillating two-state dipolar emitter close to a classically described plasmonic body\cite{novotny2012}. In that case, the time-dependent emitter's wavefunction can be expressed as $\ket{\psi(t)}= C_{g}(t)\ket{g}+C_{e}(t)\ket{e}e^{-i\omega_{eg}t}$ and so the corresponding oscillating dipole moment becomes $\bra{\psi(t)}\hat{\mu}\ket{\psi(t)}=\bra{g}\hat{\mu}\ket{g}|C_g|^2+\bra{g}\hat{\mu}\ket{e}2\Re{(C_g^*C_e)}cos(\omega_{eg}t)$ up to first-order. The oscillating contribution polarizes the nearby plasmonic body whose reaction field can act back on the dipole itself, thus leading to a self-interaction contribution mediated by the metal that is  $\propto|C_g|^2|C_e|^2f=(1-|C_e|^2)|C_e|^2f$ with \textit{f} being a complex function that accounts for the plasmon response. In the quasistatic limit the imaginary component of this self-interaction accounts for the emitter decay to the plasmonic system\cite{novotny2012} and only when $|C_e|^2 << 1$ the expression just derived reduces to the excited state population times the decay rate. In the opposite limit where the excited state population approaches 1 because of sufficiently-intense driving fields, the decay probability becomes (unphysically) zero due to the $(1-|C_e|^2)$ factor. In fact, in that limit and from a semi-classical perspective, the emitter excited state would be fully populated, and since it is a stationary solution it has no way to exchange energy with the plasmonic system as its electron density do not oscillate over time and so cannot perturb the NP. Conversely, in the full-quantum description the purely molecular state is no longer an eigenstate of the coupled Hamiltonian and so it can evolve in time to a plasmon excitation, leading to decay through the NP. This qualitative explanation is numerically verified in Fig.\,\ref{fig:high}, where the molecular excited state population upon excitation (SI 1.2, eq.19) under different driving conditions is shown. As the driving field intensity increases (panels a-c of Fig.\,\ref{fig:high}), the FQ and SC approaches start to show discrepancies in the molecular excited state population over time, and that difference is even more pronounced in the extreme limit of having the entire molecular population on the molecular excited state (panel d). In that case the SC model does not predict any form of decay, since a stationary solution of the SC system is fully populated and hence its electron density does not change over time, which prevents it from interacting with the NP and so leading to plasmon-induced decay, as discussed above. On the other hand, the FQ picture correctly captures the NP-induced molecular decay. Indeed, by fitting the FQ results with an exponential decay function a lifetime of $\approx 10\,\text{ps}$ is observed, thereby confirming the presence of an excited state decay process. In this non linear regime many quantum jumps take place (SI 1.3), resulting in the jagged profile of Fig.\,\ref{fig:high}d, which is obtained averaging over 1000 trajectories. A more formal description of this phenomenon based on fluctuation-dissipation theorem can be found in refs.\cite{cohen1986,cohen1997}.

\begin{figure}[]
    \centering
    \includegraphics[width=8.6cm]{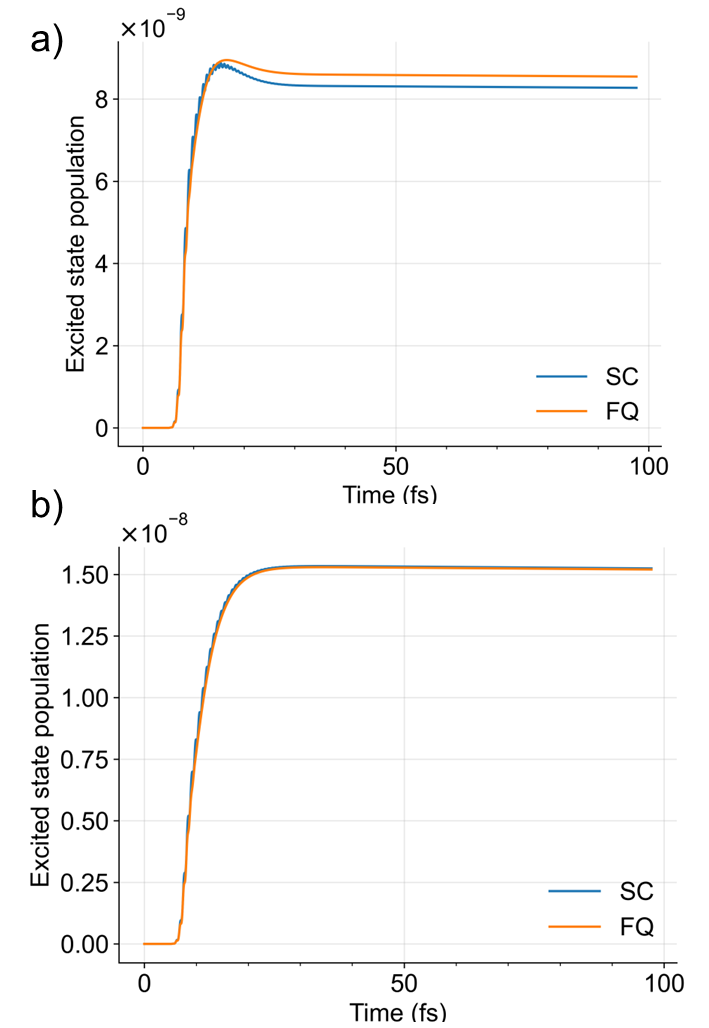}
    \caption{Molecular excited state population over time obtained via SC (blue) and FQ (orange) approaches  under different  frequency detuning  cases, (a) $\delta=\omega_e-\omega_p\approx100\,\text{meV}$ and (b)  $\delta=0$. The system (Fig.\,\ref{fig:setup}) is excited with a Gaussian pulse resonant with the lowest NP plasmon mode $\omega_p=2.95\,\text{eV} $ and whose intensity is $3.5\times10^{4}\,\text{W/cm}^2$  (see SI\,2).}
    \label{fig:multimode}
\end{figure}

On the other hand, in the weak field and weak coupling limit one could expect that both models provide the same results\cite{waks2006,waks2010}. Surprisingly,  when there is a frequency detuning between the molecular and plasmon frequencies ($\delta{\,=\,}\omega_e{\,-\,}\omega_p{\,\neq\,}0$), we observe a small-yet-appreciable difference in the results (Fig.\,\ref{fig:multimode}a). Interestingly, when the system is taken into resonance the discrepancy disappears(Fig.\,\ref{fig:multimode}b).

\noindent In order to shed light on this observed mismatch, a slightly-simplified system is considered so that an analytical model can be made to infer the origin of that discrepancy. Basically, the same setup of Fig.\ref{fig:setup} is investigated but the n. of plasmonic modes of $\hat{H}_{0,FQ}$ is restricted to one, namely only the lowest dipolar mode is considered. This approximation, which is  reasonable since the molecule-NP coupling at $\approx4\,\text{nm}$ distance is mostly dominated by dipolar interactions, is also coherently applied to the corresponding SC simulations by setting to zero the contribution to the overall classical response charges (eq.\,\ref{eq:Hsc}) originating from non-dipolar modes embedded in the NP response function\cite{pipolo2016,fregoni2021}. 

\begin{figure*}[ht!]
    \centering
    \includegraphics[width=1.0\textwidth]{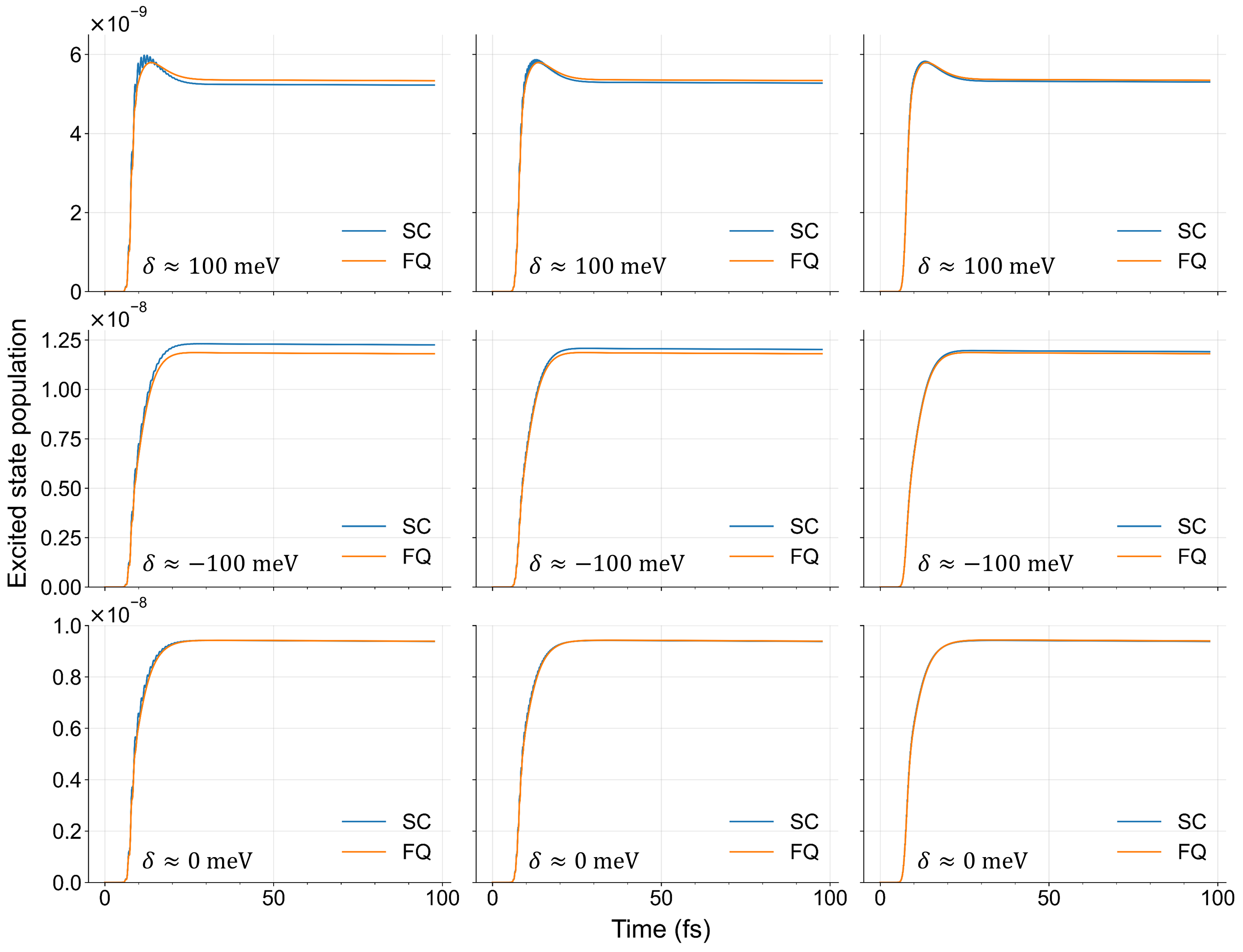}
    \caption{Molecular excited state population over time obtained via SC (blue) and FQ (orange) approaches including only the lowest plasmon dipolar mode in the models. Results under different detuning conditions are shown. Within each rank, from left to right the absolute value of the plasmon mode frequency $\omega_p$ is doubled each time and so the molecular frequency $\omega_e$ is coherently modified to preserve the same $\delta$ value among calculations displayed in the same row. Each simulation is performed by driving the system with a Gaussian pulse of same shape but resonant with the corresponding $\omega_p$ value and whose intensity is $3.5\times10^{4}\,\text{W/cm}^2$. }
    \label{fig:singlemode}
\end{figure*}

\noindent Clearly, the results displayed in Fig.\,\ref{fig:singlemode} show that when $\delta\neq0$ there is a small-yet-observable mismatch between the two approaches which vanish under resonance condition, $\delta=0$. \\ This unforeseen result can be rationalized by resorting to an analytical model for the the molecular excited state population $\left|C_{e,SC/FQ}(t)\right|^2$ in the single mode case under the weak-coupling and weak-field approximations. Indeed, using first-order perturbation theory it can be shown that (derivation is detailed in SI\,3)
\begin{widetext}
\begin{equation}\label{eq:pop}
\begin{split}
    & \left|C_{e,SC}(t)\right|^2  \approx \frac{E_{0}^2|\vec{\mu}_{e}|^2}{4} \left( 1 + \frac{4|\vec{\mu}_p|^4}{\Gamma_{p}^2|\vec{r}\,|^6}-\frac{4|\vec{\mu}_p|^4}{(4\omega_{p}^2 +\Gamma_{p}^2)|\vec{r}\,|^6}+  \frac{8|\vec{\mu}_p|^2\omega_{p}}{(4\omega_{p}^2+\Gamma_{p}^2)|\vec{r}\,|^3} \right)\frac{1}{\delta^2} \\[1ex]
    & \left|C_{e,FQ}(t)\right|^2 \approx \frac{E_0^2|\vec{\mu}_{e}|^2}{4} \left( 1+\frac{4|\vec{\mu}_p|^4}{\Gamma_p^2|\vec{r}\,|^6 }
\right)\frac{1}{\delta^2}
\end{split}
\end{equation}
\end{widetext}
\noindent where $|\vec{\mu}_{e}|$ and    $|\vec{\mu}_{p}|$ are the molecular and plasmonic transition dipoles, respectively. The latter can be easily computed as $\sum_jq_{pj}\vec{r}_{j}$ with $\vec{r}_j$ being the position vector pointing to the $j^{th}$ quantized surface charge.\\
Notably, the origin of the two additional terms appearing solely in the SC expression can be traced back to the anti-resonant term of the NP linear response polarizability (SI 3.1) which defines the classical response charges of eq.\ref{eq:Hsc} and enters in different ways in the FQ and SC models.
Indeed, if simulations are repeated scaling the Drude-Lorentz dielectric function parameters (SI 1.1, eq.2) such that the absolute value of $\omega_p$ increases while keeping the values of plasmonic charges $q_{pj}$ and detuning $\delta$ fixed (SI\,4), the observed discrepancy progressively vanish, also in the case of $\delta\neq0$. This can be understood in light of eq.\,\ref{eq:pop} where the additional terms only present in the SC expression roughly depend on $\approx \left(\omega_p\right)^{-2}$ or $\approx \left(\omega_p\right)^{-1}$, and so their contribution becomes progressively more and more negligible as the absolute value of $\omega_p$ increases, while keeping all other quantities appearing in that expression constant. This is indeed what is reported in Fig.\ref{fig:singlemode} for each value of $\delta$ moving from left to right.
Remarkably, in agreement with eq.\,\ref{eq:pop} FQ curves reported in Fig.\,\ref{fig:singlemode} do not exhibit any appreciable change as $\omega_p$ increases, whereas SC curves do vary, approaching the FQ results for large $\omega_p$ values. \\
 On the other hand, when $\delta=0$ the two curves are already almost-perfectly superimposed and nothing more can be inferred from eq.\,\ref{eq:pop} as both expressions diverge for $\delta \rightarrow 0$. Nevertheless, it can be qualitatively shown (SI 3.3) that when the molecule and plasmon excitations are resonant, the dominating contribution to the molecular excited state population is equally described numerically by both models when the driving field is resonant with the plasmon mode frequency, as in the investigated case, thus justifying, albeit not quantitatively, why under resonance condition both models yield the same result.

\section{Conclusions}\label{Conclusions}

In this work, by combining Stochastic Schr\"{o}dinger Equation, \textit{ab initio} description of target molecules and BEM-based modelling of arbitrarily-shaped plasmonic nanoparticles within the PCM-NP framework\cite{coccia2018,coccia2019,pipolo2016,fregoni2021,corni2003,mennucci2019}, we directly compare full-quantum and semi-classical modelling of plexcitons, while realistically accounting for ultrafast plasmonic dissipative processes. The presented full-quantum theory is compared with the semi-classical analogue under different excitation regimes. Upon high-intensity driving, sizeable differences between the two approaches are observed, thereby confirming previous findings on similar systems\cite{waks2006,szoke1969,rice1994,drummond1980}. Furthermore, it is surprisingly disclosed that even within the weak-field and weak-coupling limits an unexpected  small-yet-appreciable difference arises in the molecular excited state population upon low-intensity driving. By resorting to a simplified analytical model the origin of such discrepancy has been traced back to the anti-resonant term of the classical NP polarizability, which enters into the full-quantum and semi-classical models in different ways and turned out to be responsible for the observed discrepancy. The illustrated theory paves the way for real-time investigation of plexcitons beyond the linear regime while retaining \textit{ab initio} description of molecules and accurate modelling of arbitrarily-shaped plasmonic nanostructures. \\
As the field of plexcitonic chemistry is drawing ever-increasing attention,  we believe the theory presented here will be instrumental in narrowing the gap towards accurate control of molecular excited-state processes by means of plasmon-molecule coupling. \newline

\begin{acknowledgements}
The authors acknowledge Giulia Dall'Osto for insightful preliminary discussions and numerical tests. M.R. acknowledges MIUR “Dipartimenti di Eccellenza” under the project Nanochemistry for energy and Health (NExuS) for funding the Ph.D. grant. Computational work has been carried out on the C3P (Computational Chemistry Community
in Padua) HPC facility of the Department of Chemical Sciences of the University of Padua.
\end{acknowledgements}

\subsection{Supplementary Information} 
Additional material can be found in the supplementary information:\\
 Details about FQ and SC theoretical models; Computational details; Analytical models to derive eq.\,\ref{eq:pop}; Scaling plasmonic quantities to perform simulations reported in Fig.\,\ref{fig:singlemode}.

\bibliography{ref}
\end{document}


\tableofcontents
\addtocontents{toc}{\protect\thispagestyle{empty}}
\newpage

\section{Theory}\label{Theory}

\subsection{semi-classical (SC) model}\label{sub:semic}
In the semi-classical (SC) model, the molecule is described at quantum-mechanical level using standard quantum chemistry approaches, but the plasmonic NP is treated as a classical polarizable continuum object within the PCM-NP framework\cite{mennucci2019,pipolo2016}. In essence, plasmonic NPs of arbitrarily complex shape are described in the quasi-static limit, i.e. retardation effects are not included, and are coupled with a quantum chemistry description of molecules through an integral equation formalism of the Polarizable Continuum Model (IEF-PCM)\cite{mennucci2019} which basically boils down to solving the corresponding Poisson's electrostatic equation. The problem is numerically solved with a Boundary-Element-Method (BEM) approach\cite{de2002} which entails a surface discretization of the NP, leading to a set of discrete surface elements, called "tesserae", that host polarization charges due to a given external perturbation (e.g. external potential, molecular densities, etc.). Such set of charges represent the NP linear response to the external perturbation and can be used to quantitatively evaluate the molecule-NP interactions and related effects.  \\
Given these premises, the system Hamiltonian $\hat{H}_{S}(t)$ (see also SI \ref{sub:sse}), here renamed $\hat{H}_{SC}(t)$ for the actual SC case, reads
 \begin{equation}\label{eq:Hsc}
\hat{H}_{SC}(t)=\hat{H}_{mol}-\vec{\hat{\mu}}\cdot\vec{E}_{ext}(t)+(\textbf{q}_{ref}(t)+\textbf{q}_{pol}(t))\cdot\hat{\textbf{V}}
\end{equation}

\noindent where $\hat{H}_{mol}$ is the time-independent molecular Hamiltonian, $\vec{E}_{ext}(t)$ is the time-dependent external electric field that is used to drive the system, $\vec{\hat{\mu}}$ is the molecular dipole operator, $\textbf{q}_{ref}(t)$ and $\textbf{q}_{pol}(t)$ are the vectors collecting the response charges on the nanoparticle's discretized surface induced by direct polarization of the incoming exciting field ($\textbf{q}_{ref}(t)$)  and by the time-dependent nearby molecular electron density ($\textbf{q}_{pol}(t)$), and $\hat{\textbf{V}}$ is the molecular electrostatic potential operator evaluated at the nanoparticle surface where response charges lie on\cite{pipolo2016,mennucci2019}. \\ 
A time-dependent version of Boundary Element Method previously developed\cite{pipolo2016,coccia2019} (TD-BEM) is used to propagate the surface charges describing the NP response that is coupled to the molecule through the molecular electrostatic potential operator.

\noindent The nanoparticle dielectric function is described using a Drude-Lorentz (DL) model,
 \begin{equation}\label{eq:DL}
\epsilon(\omega)=1+\frac{\Omega_{plasma}^2}{\omega_0^2-\omega^2-i\Gamma\omega}
\end{equation}

\noindent where $\Omega_{plasma}$ is the metal plasma frequency, $\omega_0$ is the natural frequency of bound oscillators and $\Gamma$ is the relaxation time (damping rate) of the metal (see SI Sec.\,\ref{compdetails} for the actual values used).
\\
In the SC case, $\hat{H}_{mol}$ of eq.\ref{eq:Hsc} is purely the molecular Hamiltonian, so the full system wavefunction, $\ket{\psi_{S}(t)}$, here renamed as $\ket{\psi_{SC}(t)}$, can be expanded on the basis of the molecular stationary eigenstates $\ket{m}$ as
 \begin{equation}\label{eq:exp}
\ket{\psi_{SC}(t)}=\sum_{m}^{N_{states}}C_{m}(t)\ket{m}
\end{equation}

\noindent where the molecular eigenstates $\ket{m}$ can be obtained with any quantum chemistry approach. In our case we use Configuration Interaction Singles (CIS, SI\,\ref{compdetails}) and the resulting Stochastic Schr\"{o}dinger Equation (see SI \ref{sub:sse}) we aim to solve in matrix form reads:
 \begin{equation}\label{eq:sse_m}
i\frac{\partial \boldsymbol{C}(t)}{\partial t}=\boldsymbol{H}_{SSE,SC}(t)\boldsymbol{C}(t)
\end{equation}
\noindent with 
 \begin{equation}\label{eq:Hsse_sc}
\hat{H}_{SSE,SC}(t)=\hat{H}_{SC}(t)-\frac{i}{2}\sum_{q}^{M}\hat{S}_{q,SC}^{\dagger}\hat{S}_{q,SC}\ .
\end{equation}

\noindent $\boldsymbol{C}(t)$ is the vector of the time-dependent coefficients of eq.\ref{eq:exp} describing the wavefunction at a given time step \textit{t} represented on the basis of the molecular eigenstates, whereas $\hat{H}_{SSE,SC}(t)$ is the time-dependent Hamiltonian that is used for propagating the system wavefunction according to a second-order Euler algorithm\cite{coccia2018} (see also SI \ref{sub:sse}). \\
The dissipative operators $\hat{S}_{q,SC}$ in eq.\ref{eq:Hsse_sc} are still to be defined. Indeed, different choices may be taken, depending on the relevant decay processes that are considered for a given system. Herein, since the main goal of the present work is to compare the SC and FQ models on a perfectly consistent ground, we solely focus on the NP-induced non radiative decay, which is intrinsically included in $\hat{H}_{SC}(t)$ through $\textbf{q}_{pol}(t)$, as mentioned in main text.   \\
Under such assumptions eq.\ref{eq:Hsse_sc} simplifies to 
 \begin{equation}\label{eq:Hsse_sc2}
\hat{H}_{SSE,SC}(t)=\hat{H}_{SC}(t).
\end{equation}

\subsection{full-quantum (FQ) model}\label{sub:fullq}

In the full-quantum picture (FQ), the plasmonic NP is also quantized, so the system Hamiltonian becomes
\begin{equation}\label{eq:Hfq}
    \hat{H}_{FQ}(t)=\hat{H}_{0,FQ}-\vec{\hat{\mu}}\cdot\vec{E}_{ext}(t)
\end{equation}

\noindent where $\hat{H}_{0,FQ}$  is the full plasmon-molecule Hamiltonian\cite{fregoni2021},
\begin{equation}\label{eq:Hfq_full}
    \hat{H}_{0,FQ}= \hat{H}_{mol}+\sum_p\omega_{p}{\hat{b}}^{\dagger}_{p}{\hat{b}}_{p}+\sum_{pj}q_{pj}\hat{V}_{j}({b}^{\dagger}_{p}+{b}_{p})
\end{equation}

\noindent with $\omega_p$ being the frequency of the $p^{th}$ quantized plasmon mode and $q_{pj}$ being the corresponding quantized surface charge lying on the $j^{th}$ tessera. $\hat{b}^{\dagger}_{p}$ and $\hat{b}_p$ are the corresponding plasmonic creation and annhilation operators and $\hat{V}_{j}$ is instead the molecular electrostatic potential operator evaluated at the $j^{th}$ tessera. The complete derivation of the Q-PCM-NP quantization scheme that leads to eq.\ref{eq:Hfq_full} is reported in ref.\cite{fregoni2021}. \\
Starting from eq.\ref{eq:Hfq_full} and assuming to consider two molecular states only $\ket{g},\ket{e}$, the Hamiltonian of eq.\ref{eq:Hfq_full} can be recast into a more familiar form after inserting the molecular identity operator $\mathbb{1_{mol}}=\ket{g}\bra{g}+\ket{e}\bra{e}$ before and after $\hat{V}_j$, leading to
\begin{equation}
\begin{split}\label{eq:Hfq_full2}
    \hat{H}_{0,FQ}= &
    \omega_g\ket{g}\bra{g}+\omega_e\ket{e}\bra{e}+\sum_p\omega_{p}{\hat{b}}^{\dagger}_{p}{\hat{b}}_{p}+\sum_{pj}q_{pj}\left(\hat{V}_{j}^{gg}\ket{g}\bra{g}+\hat{V}_{j}^{ee}\ket{e}\bra{e}+ \right. \\ 
    & \left.\hat{V}_{j}^{eg}\ket{e}\bra{g}+\hat{V}_{j}^{ge}\ket{g}\bra{e}\right)\left({b}^{\dagger}_{p}+{b}_{p}\right)
    \end{split}
\end{equation}

\noindent where $\omega_g$ and $\omega_e$ are the energies of the corresponding molecular states and the shorthand notation $\hat{V}_j^{eg}$ stands for $\bra{e}\hat{V}_j\ket{g}$.The diagonal terms of the plasmon-molecule interaction $\sum_{pj}q_{pj}\hat{V}_{j}^{gg},\,\sum_{pj}q_{pj}\hat{V}_{j}^{ee}$ in eq.\ref{eq:Hfq_full2} lead to a correction of the molecular excitation frequency $\omega_e-\omega_g$ due to the nearby NP. This contribution is numerically negligible in the present case, therefore upon setting $\omega_g=0$ so that $\omega_e$ becomes the molecular transition frequency, we end up with the simplified expression
\begin{equation}\label{eq:Hfq_full3}
 \hat{H}_{0,FQ}= 
    \omega_{e}{\hat{\sigma}}^{\dagger}{\hat{\sigma}}+\sum_p\omega_{p}{\hat{b}}^{\dagger}_{p}{\hat{b}}_{p}+\sum_{pj}q_{pj}\hat{V}_{j}^{eg}
  \left(\hat{\sigma}^{\dagger}+\hat{\sigma}\right)\left({b}^{\dagger}_{p}+{b}_{p}\right)
\end{equation}
where we have introduced the molecular transfer operators $\hat{\sigma}^{\dagger}=\ket{e}\bra{g}$, $\hat{\sigma}=\ket{g}\bra{e}$ and $\hat{V}_{j}^{eg}$ can be taken to be real.
Clearly, eq.\ref{eq:Hfq_full3} leads to the well-known Jaynes-Cummings multimode Hamiltonian\cite{jaynes1963,romanelli2023} after applying the usual rotating wave approximation to the interaction terms, namely 
\begin{equation}\label{eq:Hfq2}
    \hat{H}_{0,FQ}= \omega_{e}{\hat{\sigma}}^{\dagger}{\hat{\sigma}}+\sum_p\omega_{p}{\hat{b}}^{\dagger}_{p}{\hat{b}}_{p}+\sum_pg_{p}({\hat{b}}^{\dagger}_{p}{\hat{\sigma}}+{\hat{b}}_{p}{\hat{\sigma}}^{\dagger})
\end{equation}

\noindent with $g_p$ being the corresponding  transition coupling element computed as,
\begin{equation}
g_{p}=\bra{e,0}\sum_{j}q_{pj}{\hat{V}}_{j}({b}^{\dagger}_{p}+{b}_{p})\ket{g,1_{p}}
\label{eq:gp}
\end{equation}
 in the single-excitation subspace. 
In eq.\ref{eq:gp} the shorthand notation $\ket{e,0},\ket{g,1_p}$ has been introduced which represents the molecular excited state with all plasmon modes in their ground state and the $p^{th}$ plasmon mode singly excited with the molecule in its ground state, respectively. Both approximations that have been applied moving from eq.\ref{eq:Hfq_full2} to eq.\ref{eq:Hfq2} were numerically tested in the investigated case and they proved to be valid .

The Hamiltonian of eq.\ref{eq:Hfq2}, expressed in the 1-excitation states basis\cite{delpo2020,kuttruff2022}, can be fully diagonalized to obtain the 1-excitation plexcitonic states ($\ket{\widetilde{m}}$), which can be generally expressed as
\begin{equation}\label{eq:psi_pl}
\ket{\widetilde{m}}= \sum_p\ket{e,0}C^{\hspace{0.15mm}\widetilde{m}}_{e}+ \ket{g,1_p}C^{\hspace{0.15mm}\widetilde{m}}_p.
\end{equation}

\noindent In the FQ picture, the complete Hamiltonian that is used for time-propagation  becomes
 \begin{equation}\label{eq:Hsse_fq}
\hat{H}_{SSE,FQ}(t)=\hat{H}_{FQ}(t)-\frac{i}{2}\sum_p\hat{S}_{p,FQ}^{\dagger}\hat{S}_{p,FQ}
\end{equation}
\noindent with
 \begin{equation}\label{eq:S_fq}
\hat{S}_{p,FQ}=\sqrt{\Gamma_p}\mathbb{1_{mol}}\otimes\left(\ket{0}\bra{1_p}\right)
\end{equation}
\noindent where $\Gamma_p$ is the decay rate of the $p^{th}$ quantized mode that is actually equal to $\Gamma$ (eq.\ref{eq:DL}) for every $p^{th}$ plasmon mode originating from the Q-PCM-NP quantization scheme\cite{fregoni2021}. \\
Clearly, the form of the chosen decay operator of eq.\ref{eq:S_fq} is an assumption which implies that the only  source of dissipation in the current FQ picture comes from the plasmonic part of the wavefunction. This approximation basically neglects any other intrinsic molecular relaxation channel, which often take place on a longer time scale (tens/hundreds of picoseconds). The same assumption has been consistently made in the SC model where only the NP-mediated non radiative decay has been included (see SI \ref{sub:semic}).

 In the FQ picture the time-dependent wavefunction can be generally expressed in the plexcitonic basis (eq.\ref{eq:psi_pl})  as 
\begin{equation}\label{eq:exp_fq}
\ket{\psi_{FQ}(t)}=\sum_{\widetilde{m}}C_{\widetilde{m}}(t)\ket{\widetilde{m}},
\end{equation}

\noindent thus leading to the following equation 
 \begin{equation}\label{eq:sse_m_fq}
i\frac{\partial \boldsymbol{C(t)}}{\partial t}=\boldsymbol{H}_{SSE,FQ}(t)\boldsymbol{C(t)}
\end{equation}
\noindent which is the FQ analogue to eq.\ref{eq:sse_m}, where now $\boldsymbol{C(t)}$ is the vector of time-dependent coefficients representing the wavefunction on the plexcitonic basis. 

\noindent On this basis  $\hat{H}_{0,FQ}$ is diagonal, but $\hat{S}_{p,FQ}^{\dagger}\hat{S}_{p,FQ}$ is not. Indeed, using eqs.\ref{eq:psi_pl}-\ref{eq:S_fq} it is easy to show that
 \begin{equation}\label{eq:S_fq2}
 \begin{split}
 & \bra{\widetilde{m}}\hat{S}_{p,FQ}^{\dagger}\hat{S}_{p,FQ}\ket{\widetilde{m}}=\left(C^{\hspace{0.15mm}\widetilde{m}}_p\right)^2\Gamma_p   \\
 & \bra{\widetilde{n}}\hat{S}_{p,FQ}^{\dagger}\hat{S}_{p,FQ}\ket{\widetilde{m}}=\left(C^{\hspace{0.15mm}\widetilde{n}}_pC^{\hspace{0.15mm}\widetilde{m}}_p\right)\Gamma_p 
\end{split}
\end{equation}

\noindent where the coefficients $C^{\hspace{0.15mm}\widetilde{n}/\widetilde{m}}_p$ can be taken to be real because of the form of $\hat{H}_{0,FQ}$ (eq.\ref{eq:Hfq2}).

\indent The physical quantity that is investigated in the present work to compare SC and FQ descriptions (see Figs. 2-4, main text) is the molecular excited state population upon driving, which according to eq.\ref{eq:exp} and eqs.\ref{eq:psi_pl},\ref{eq:exp_fq} respectively reads
\begin{equation}
\begin{split}
& \left|C_{e,SC}(t)\right|^2=\left|\braket{e|\psi_{SC}(t)}\right|^2=\left| C_e(t) \right|^2\ , \\
&\left|C_{e,FQ}(t)\right|^2=\left|\braket{e,0|\psi_{FQ}(t)}\right|^2=\left| \sum_{\widetilde{m}}C^{\hspace{0.15mm}\widetilde{m}}_{e}C_{\widetilde{m}}(t)\right|^2.
\label{eq:pop0}
\end{split}
\end{equation}

\subsection{Stochastic Schr\"{o}dinger Equation}\label{sub:sse}
In the Markovian limit, an open-quantum-system description of the system-bath interaction leads to the following Stochastic Schr\"{o}dinger equation (SSE) expressed in atomic units (\text{a.u.})
 \begin{equation}\label{eq:sse}
i\frac{d}{dt}\ket{\psi_{S}(t)}=\hat{H}_{S}(t)\ket{\psi_{S}(t)}+\sum_{q}^{M}l_{q}(t)\hat{S}_{q}\ket{\psi_{S}(t)}-\frac{i}{2}\sum_{q}^{M}\hat{S}_{q}^{\dagger}\hat{S}_{q}\ket{\psi_{S}(t)}.
\end{equation}

\noindent $\ket{\psi_{S}(t)}$ and $\hat{H}_{S}(t)$ are the system time-dependent wavefunction and  Hamiltonian, respectively, whose definitions depend on the model being used for describing the nanoparticle-molecule system (see SI \ref{sub:semic} and \ref{sub:fullq} ). The operators $\hat{S}_{q}$ describe the effect of the surroundings (bath) on the system through different M interaction channels, each one labelled as \textit{q} and defined according to the type of dissipative process that is modelled (e.g. non-radiative decay, dephasing etc.). The non-Hermitian term of eq.\ref{eq:sse} , $-\frac{i}{2}\sum_{q}^{M}\hat{S}_{q}^{\dagger}\hat{S}_{q}$, represents dissipation due to the environment, while $\sum_{q}^{M}l_{q}(t)\hat{S}_{q}$ is a fluctuation term modelled as a Wiener process $l_{q}(t)$ i.e. white noise associated with the Markovian approximation. In our case, the SSE (eq.\ref{eq:sse}) is propagated using a quantum jump algorithm which practically translates to accounting for the explicit fluctuation term of eq.\ref{eq:sse} through a Monte-Carlo like method based on random quantum jumps\cite{dalibard1992,molmer1993,dum1992,coccia2018}. We remark that the explicit form of the operators appearing in eq.\ref{eq:sse} depend on the model that is used for describing the coupled system, which then differs between the SC and FQ approaches described above. 
Since this is a stochastic process, an independent number of trajectories $N_{traj}$ have to be performed and by averaging the corresponding results, system properties, like populations and coherences for instance, can be obtained. In the ideal limit of an infinite number of independent realizations, the averaged results match those coming from time-propagating the system reduced density-matrix with a Lindblad-like master equation approach\cite{coccia2018}.

\section{Computational details}\label{compdetails}

The quinolone molecule is described at \textit{ab initio} level, using CIS/6-31g* in a locally-modified version of GAMESS\cite{pipolo2016,schmidt1993} which accounts for the presence of the nearby NP in the determination of the molecular ground state at the classical level. The molecular excited states are determined assuming the NP classical polarization remains frozen to that proper for ground state. This means that for both the SC and FQ models, the same set of states are used. Moreover, the use of the rotating wave approximation (see eq.\ref{eq:Hfq2}) in the FQ model does not provide further change to the ground state.

According to adopted level of theory, the lowest bright excited state of quinolone features an excitation energy of $\approx 2.95\,\text{eV}$.
The NP optical response is modelled with a Drude-Lorentz dielectric function (eq.\ref{eq:DL}) setting $\Omega_{plasma}=0.240\,\text{a.u.} \approx 6.5 \,\text{eV}$, which is close to values previously adopted for gold\cite{zeman1987}, $\omega_0=0\,\text{a.u.}$ and $\Gamma=0.01515\,\text{a.u.}$, that corresponds to a lifetime of $\approx 2\,\text{fs}$. With such parameters, the lowest dipolar plasmon mode of the NP is basically in resonance with the molecular transition ($\omega_p\approx2.95\,\text{eV}$).The computed transition coupling element (eq.\ref{eq:gp}) between that mode, which is the most relevant since all others lie higher in energy and are scarcely coupled with the molecule, and the lowest quinolone bright excitation is $|g| \approx 2.5\,\text{meV}$ which, given the value of $\Gamma=0.01515\,\text{a.u.}\approx410\,\text{meV}$, results in $\frac{|g|}{\Gamma}<<1$, thus setting the present case in the weak-coupling regime\cite{dovzhenko2018}. \\
Time-dependent simulations have been performed by exciting the system with a pulse with gaussian envelope, resonant with the dipolar plasmon mode frequency $\omega_p=2.95\,\text{eV}$,
 \begin{equation}\label{eq:field_f}
\vec{E}_{ext}(t)=\hat{r}E_{0}exp\left( -\frac{\left(t-t_0\right)^2}{2\sigma^2}\right)cos(\omega_pt)
\end{equation}
where $\hat{r}$ is the unit vector pointing along the molecule-NP distance direction (Fig.1, main text), $E_0$ is the field amplitude set to $10^{-6} \text{a.u.}$ unless differently specified ($\approx5\times10^5\,\text{V/m}$, corresponding to $ 3.5\times10^4\,\text{W/cm}^2$ light intensity) which ensures the weak-field limit\cite{boyd2020}, $t_0\approx7\,\text{fs}$ and $\sigma\approx1\,\text{fs}$. The coefficients $\boldsymbol{C(t)}$ of   eqs.\ref{eq:sse_m},\ref{eq:sse_m_fq} are propagated via a second-order Euler algorithm in combination with quantum jumps\cite{coccia2018,coccia2019} using a time step of $0.1\,\text{as}$. Under weak-field driving 100 trajectories were run initially, but no quantum jump occurred under weak-field excitation, so results shown in Figs. 3-4 (main text) actually originate from individual wavefunction propagation. On the other hand, beyond the weak-field limit (Fig.2 main text) quantum jumps do often occur in FQ simulations, and so averaging over 1000 trajectories has been done in that case.

\section{Analytical models}
\subsection{semi-classical picture in the weak-coupling limit}\label{sec:SC}
Starting from the full SC Hamiltonian of eq.\ref{eq:Hsc} and considering that the molecule-NP distance of the investigated setup (Fig.1 main text) is large enough to ensure that only dipolar interactions are relvant,  we can apply the dipole approximation to the molecule-NP coupling terms (rightmost terms of eq.\ref{eq:Hsc}, SI \ref{sub:semic}). Additionally, the $\textbf{q}_{pol}(t)$ contribution to the classical response charges of eq.\ref{eq:Hsc} is typically much smaller than $\textbf{q}_{ref}(t)$ when the plasmonic system is excited nearby plasmon resonance, the latter being directly related to the scattered field of the NP due to external driving, and so it can be neglected, thus leading to 
\begin{equation}\label{eq:Hsc_dip}
    \hat{H}_{SC}(t)=\hat{H}_{mol}-\vec{\hat{\mu}}\cdot\vec{E}_{ext}(t)-\vec{\hat{\mu}}\cdot\vec{E}_{scatt}(t)
\end{equation}

\noindent where $\vec{E}_{scatt}(t)$ is the scattered field of the NP at the molecule location upon external excitation. In other words, this contribution accounts for the plasmonic local field enhancement effect.

\begin{figure}[ht!]
    \centering
    \includegraphics[width=1.0\textwidth]{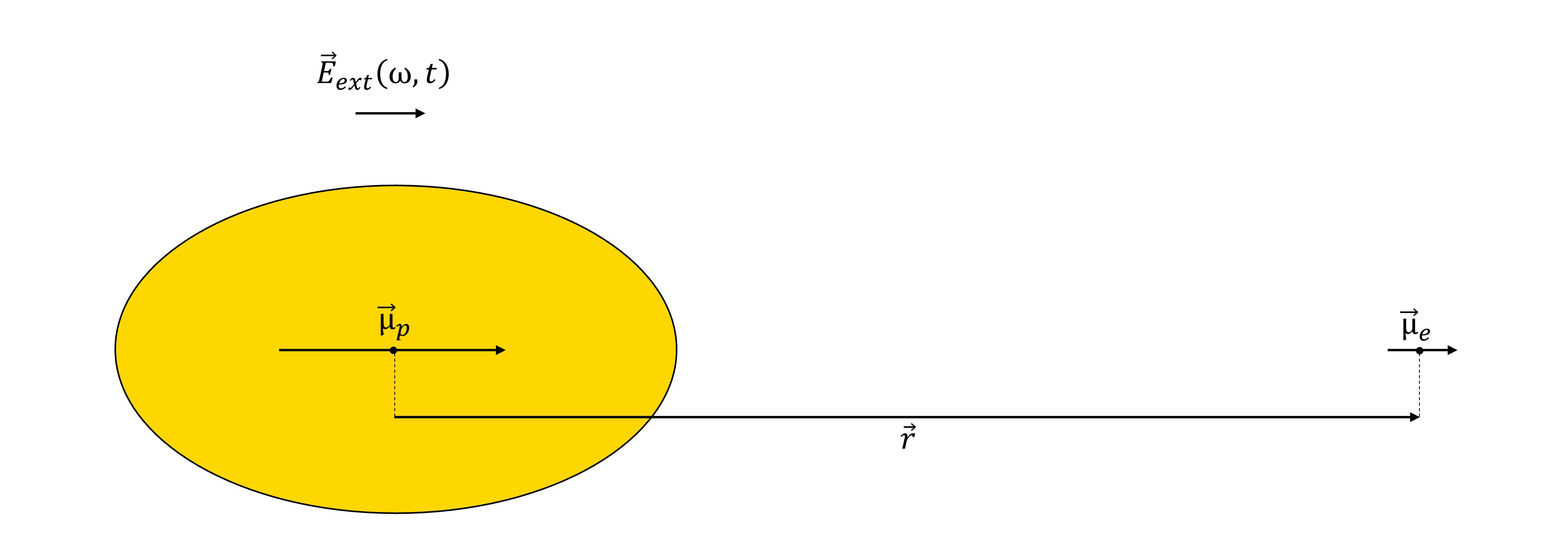}
    \caption{Schematic representation of the NP-molecule setup considered for the analytical model. $\vec{\mu}_p$ and $\vec{\mu}_{e}$ represent the plasmonic and molecular transition dipoles, respectively.}
    \label{fig:setup}
\end{figure}

\noindent The setup that has been investigated (Fig.1 main text) features a large molecule-NP distance which basically ensures the validity of the weak-coupling condition. Under these circumstances, the scattered field of the NP at the molecule position is dominated by the plasmonic dipolar response, which can then be formally expressed as
\begin{equation}\label{eq:alfa1}
    \vec{E}_{scatt}(\omega,\vec{r})=
    \frac{3\left[ \left( \vec{E}_{ext}(\omega)\alpha_{NP}(\omega)\right)\cdot{\hat{r}}\right]{\hat{r}}-\vec{E}_{ext}(\omega)\alpha_{NP}(\omega)}{|\vec{r}\,|^3}
\end{equation}

\noindent with $\hat{r}$ being the unit vector pointing along the NP-molecule direction (Fig.\ref{fig:setup}) and $\alpha_{NP}(\omega)$ being the frequency-dependent NP polarizability expressed as\cite{fregoni2021}
\begin{equation}\label{eq:alfa2}
    \alpha_{NP}(\omega)=\sum_{p}\frac{\bra{0}\hat{\mu}\ket{p}\bra{p}\hat{\mu}\ket{0}}{\omega_{p}-\omega-i\Gamma_{p}} +
    \frac{\bra{p}\hat{\mu}\ket{0}\bra{0}\hat{\mu}\ket{p}}{\omega_{p}+\omega+i\Gamma_{p}}
\end{equation}

\noindent where $\omega_p$ and $\Gamma_p$ are the frequency and damping rate of the corresponding $p^{th}$ plasmonic mode. \\
In principle the sum of eq.\ref{eq:alfa2} should run over all the plasmon modes of the NP,
however, in the investigated weak-coupling limit the NP-molecule distance is large enough such that only dipolar resonances significantly contribute to the scattered field at the molecule location, thus eq.\ref{eq:alfa2} can be reasonably approximated to
\begin{equation}\label{eq:alfa3}
    \alpha_{NP}(\omega)\approx\frac{|\vec{\mu}_p|^2}{\omega_{p}-\omega-i\Gamma_{p}} +
    \frac{|\vec{\mu}_p|^2}{\omega_{p}+\omega+i\Gamma_{p}}
\end{equation}

\noindent with $|\vec{\mu}_p|^2$ being the squared modulus of the transition dipole moment of the only-relevant dipolar plasmonic mode. 
\\
In our model, the NP dipolar mode is exactly aligned with the direction of the incoming external field and the molecular transition dipole (Fig.\ref{fig:setup}), thus eq.\ref{eq:alfa1} simplifies to
\begin{equation}
\label{eq:Escatt}
\vec{E}_{scatt}(\omega,\vec{r})=\frac{2\vec{E}_{ext}(\omega)\alpha_{NP}(\omega)}{|\vec{r}\,|^3}
\end{equation}
and the Hamiltonian of eq.\ref{eq:Hsc_dip} becomes
\begin{equation}\label{eq:Hsc_dip2}
    \hat{H}_{SC}(t)=\hat{H}_{mol}-\vec{\hat{\mu}}\cdot\vec{E}_{ext}(t)\left( 1+\frac{2\alpha_{NP}(\omega)}{|\vec{r}\,|^3}\right).
\end{equation}

\noindent Given these premises, the excited state molecular population plotted in Fig.4 main text can be inferred by means of time-dependent perturbation theory, where the second (time-dependent) term of the r.h.s of eq.\ref{eq:Hsc_dip2} is the actual perturbation $\hat{V}(t)$. First-order time-dependent perturbation theory leads to the usual expression for excited state coefficients\cite{atkins2011}, namely
\begin{equation}\label{eq:ce1}
\begin{split}
    C_{e,SC}^{'}(t) & =-i\int_{-\infty}^{t}\bra{e}\hat{V}(t)\ket{g}e^{i\omega_{e}t}dt \\
    & =i\int_{-\infty}^{t}\bra{e}\hat{\mu}\ket{g}\cdot\vec{E}_{ext}(t) \left( 1 + \frac{2|\vec{\mu}_p|^2}{(\omega_{p}-\omega-i\Gamma_{p})|\vec{r}\,|^3} +
    \frac{2|\vec{\mu}_p|^2}{(\omega_{p}+\omega+i\Gamma_{p})|\vec{r}\,|^3} \right) e^{i\omega_{e}t}dt
\end{split}
\end{equation}

\noindent where $\omega_e$ is the excitation frequency of the considered molecular transition. \\
Assuming then the external driving is a monochromatic oscillating perturbation in resonance with the plasmon frequency and oriented along the transition dipoles (Fig.\ref{fig:setup}), that is
\begin{equation}\label{eq:field}
\vec{E}_{ext}(\omega_{p})=\hat{r}E_{0}\frac{(e^{i\omega_{p}t}+e^{-i\omega_{p}t})}{2}
\end{equation}
\noindent we end up with
\begin{equation}\label{eq:ce2}
    C_{e,SC}^{'}(t) =i\frac{E_{0}|\vec{\mu}_{e}|}{2}\left( 1 + \frac{2|\vec{\mu}_p|^2}{(-i\Gamma_{p})|\vec{r}\,|^3} +
    \frac{2|\vec{\mu}_p|^2}{(2\omega_{p}+i\Gamma_{p})|\vec{r}\,|^3} \right) \int_{-\infty}^{t} e^{i(\omega_{e}-\omega_{p})t}+e^{i(\omega_{e}+\omega_{p})t}dt
\end{equation}

\noindent where $|\vec{\mu}_{e}|$ is the molecular transition dipole associated to the transition between the ground and the first excited state ($\ket{g}\rightarrow \ket{e}$).
Solving\footnote{To be formally correct one should consider a slowly switching perturbation as $\lim_{\epsilon\to0} e^{\epsilon t}\hat{V}(t)$ which basically enable us to set the value of the integral at the lower limit to zero. This has been implicitly done passing from eq.\ref{eq:ce2} to eq.\ref{eq:ce3}} the integral of eq.\ref{eq:ce2} and rearranging some terms results in
\begin{equation}\label{eq:ce3}
\begin{split}
    C_{e,SC}^{'}(t) & =\frac{E_{0}|\vec{\mu}_{e}|}{2}\left( 1 + \frac{i2|\vec{\mu}_p|^2}{\Gamma_{p}|\vec{r}\,|^3} +
    \frac{2|\vec{\mu}_p|^2(2\omega_{p}-i\Gamma_{p})}{(4\omega_{p}^2+\Gamma_{p}^2)|\vec{r}\,|^3} \right)\cdot \left( \frac{e^{i(\omega_{e}-\omega_{p})t}}{\omega_{e}-\omega_{p}}+ \frac{e^{i(\omega_{e}+\omega_{p})t}}{\omega_{e}+\omega_{p}} \right) \\
    & =\frac{E_{0}|\vec{\mu}_{e}|}{2}\left( 1 + \frac{i2|\vec{\mu}_p|^2}{\Gamma_{p}|\vec{r}\,|^3} +
    \frac{2|\vec{\mu}_p|^2(2\omega_{p}-i\Gamma_{p})}{(4\omega_{p}^2+\Gamma_{p}^2)|\vec{r}\,|^3} \right)\cdot \left( \frac{e^{i\delta t}}{\delta}+ \frac{e^{i(2\omega_{p}+\delta)t}}{2\omega_{p}+\delta} \right) 
    \end{split}
\end{equation}

\noindent where $\delta=\omega_{e}-\omega_{p}$, which is the frequency detuning between the molecular and plasmon frequencies, has been introduced for convenience. \\
Finally, taking the squared modulus of eq.\ref{eq:ce3} leads to the $1^{st}$-order expression of the molecular excited state population, that is
\begin{equation}\label{eq:pop}
\begin{split}
    |C_{e,SC}^{'}(t)|^2  =\frac{E_{0}^2|\vec{\mu}_{e}|^2}{4} & \left ( 1 + \frac{16|\vec{\mu}_p|^4\omega_{p}^2}{(4\omega_{p}^2 +\Gamma_{p}^2)^{2}|\vec{r}\,|^6}+ 
    \frac{8|\vec{\mu}_p|^2\omega_{p}}{(4\omega_{p}^2+\Gamma_{p}^2)|\vec{r}\,|^3} +
    \frac{4|\vec{\mu}_p|^4}{\Gamma_{p}^2|\vec{r}\,|^6}+
     \frac{4|\vec{\mu}_p|^4\Gamma_{p}^2}{(4\omega_{p}^2+\Gamma_{p}^2)^2|\vec{r}\,|^6} \right. \\
     & \left. -\frac{8|\vec{\mu}_p|^4}{(4\omega_{p}^2+\Gamma_{p}^2)|\vec{r}\,|^6} \right) \cdot
      \left( \frac{1}{\delta^2}+ \frac{1}{(2\omega_{p}+\delta)^2}+2\mathfrak{Re}\Bigl\{\frac{e^{-i2\omega_{p}t}}{(2\omega_{p}+\delta)\delta}\Bigl\}   \right) 
\end{split}
\end{equation}
\noindent which simplifies to 
\begin{equation}\label{eq:pop2}
\begin{split}
    |C_{e,SC}^{'}(t)|^2  =\frac{E_{0}^2|\vec{\mu}_{e}|^2}{4} & \left( 1 + \frac{4|\vec{\mu}_p|^4}{\Gamma_{p}^2|\vec{r}\,|^6} -\frac{4|\vec{\mu}_p|^4}{(4\omega_{p}^2 +\Gamma_{p}^2)|\vec{r}\,|^6}+
     \frac{8|\vec{\mu}_p|^2\omega_{p}}{(4\omega_{p}^2+\Gamma_{p}^2)|\vec{r}\,|^3} \right) \cdot \\
      & \left( \frac{1}{\delta^2}+ \frac{1}{(2\omega_{p}+\delta)^2} +  2\mathfrak{Re}\Bigl\{\frac{e^{-i2\omega_{p}t}}{(2\omega_{p}+\delta)\delta}\Bigl\}   \right)\ .
\end{split}
\end{equation}

\noindent Eq.\ref{eq:pop2} is the main result of the SC derivation that is compared in the following to the analogous quantity obtained from the FQ model (SI \ref{sec:FQ}, below).

\subsection{full-quantum picture in the weak-coupling limit}\label{sec:FQ}

In the FQ model, the Hamiltonian of interest is
\begin{equation}
    \hat{H}_{FQ}(t)=\hat{H}_{0,FQ}-\vec{\hat{\mu}}\cdot\vec{E}_{ext}(t)
\end{equation}
and $\hat{H}_{0,FQ}$ (SI \ref{sub:fullq}) in the single-mode case reads
\begin{equation}\label{eq:Hfq2_single}
    \hat{H}_{0,FQ}= \omega_{e}{\sigma}^{\dagger}{\sigma}+\omega_{p}{b}^{\dagger}_{p}{b}_{p}+g_{p}({b}^{\dagger}_{p}{\sigma}+{b}_{p}{\sigma}^{\dagger}).
\end{equation}

\noindent As explained in SI \ref{sub:fullq}, the molecule-NP interaction in the FQ picture enters directly into the $0^{th}$ order time-independent Hamiltonian, and the remaining time-dependent perturbation is just the interaction with the external field. \\ Diagonalization of eq.\ref{eq:Hfq2_single} in the 1-excitation states manifold leads to the usual expression of plexcitonic wavefunctions
\begin{equation}\label{eq:psi}
\begin{split}
&\ket{LP}= \ket{e,0}C^{LP}_{e}+\ket{g,1}C^{LP}_p    \\
&\ket{UP}= \ket{e,0}C^{UP}_{e}+\ket{g,1}C^{UP}_{p}.
\end{split}   
\end{equation}

\noindent The first-order wavefunction upon interaction with the external field can be expressed as\cite{atkins2011}
\begin{equation}\label{eq:psi_fq}
\ket{\psi^{'}_{FQ}(t)}= C_{g}^{'}(t)\ket{g,0}e^{-i\omega_{g}t}+C_{LP}^{'}(t)\ket{LP}e^{-i\omega_{LP}t}+C_{UP}^{'}(t)\ket{UP}e^{-i\omega_{UP}t}
\end{equation}

 \noindent and since the focus of the present derivation is to obtain an analytical expression for the molecular excited state population so as to compare with the SC analogue (eq.\ref{eq:pop2}), we have 
\begin{equation}\label{eq:ce_fq}
C_{e,FQ}^{'}(t)= \braket{e,0|\psi^{'}_{FQ}(t)}=
C_{LP}^{'}(t)C^{LP}_{e}\ket{LP}e^{-i\omega_{LP}t}+C_{UP}^{'}(t)C^{UP}_{e}\ket{UP}e^{-i\omega_{UP}t}.
\end{equation}

\noindent Time-dependent pertubation theory can be again used to obtain an expression for first-order coefficients, namely
\begin{equation}\label{eq:cfq}
\begin{split}
    & C_{UP}^{'}(t)=i\frac{E_0}{2}\int_{-\infty}^{t}\bra{UP}\hat{\mu}\ket{g,0}\cdot\hat{r}
    \left(e^{i(\omega_{UP}-\omega_{p})t}+e^{i(\omega_{UP}+\omega_{p})t}\right)dt \\
    & C_{LP}^{'}(t)=i\frac{E_0}{2}\int_{-\infty}^{t}\bra{LP}\hat{\mu}\ket{g,0}\cdot\hat{r}
    \left(e^{i(\omega_{LP}-\omega_{p})t}+e^{i(\omega_{LP}+\omega_{p})t}\right)dt
\end{split}
\end{equation}

\noindent where the same monochromatic field as in the semi-classical case (eq.\ref{eq:field}) has been used.  \\
Given the relations of eqs.\ref{eq:psi},\ref{eq:cfq} and considering that the electric field direction ($\hat{r}$) is aligned with both the molecular and plasmonic transition dipoles, substitution of eq.\ref{eq:cfq} into eq.\ref{eq:ce_fq} upon solving the integrals leads to
\begin{equation}\label{eq:ce_fq2}
\begin{split}
C_{e,FQ}^{'}(t)= & \frac{E_0}{2}\left(\left|C^{UP}_{e}\right|^2\left|\vec{\mu}_{e}\right|+C^{UP}_{e}
\left(C^{UP}_{p}\right)^*\left|\vec{\mu}_p\right|\right)\left( \frac{e^{-i\omega_{p}t}}{\omega_{UP}-\omega_{p}}+ \frac{e^{i\omega_{p}t}}{\omega_{UP}+\omega_{p}}
\right) + \\
&\frac{E_0}{2}\left(\left|C^{LP}_{e}\right|^2\left|\vec{\mu}_{e}\right|+C^{LP}_{e}
\left(C^{LP}_{p}\right)^*\left|\vec{\mu}_p\right|\right)\left( \frac{e^{-i\omega_{p}t}}{\omega_{LP}-\omega_{p}}+ \frac{e^{i\omega_{p}t}}{\omega_{LP}+\omega_{p}}
\right).
\end{split}
\end{equation}

\noindent Eq.\ref{eq:ce_fq2} can be further manipulated to obtain an expression more similar to the SC analogue. Indeed, since we are in the weak-coupling limit (with $\delta=\omega_{e}-\omega_{p}>0$) time-independent perturbation theory can be used to obtain an approximate quantitative expressions for $C^{UP/LP}_{e}$ and $\omega_{UP/LP}$, resulting in\cite{atkins2011}
\begin{equation}\label{eq:tipt}
\begin{split}
&\omega_{UP}\approx\omega_{e}+\frac{g^2}{\tilde{\delta}} \\
&\omega_{LP}\approx\overline{\omega_{p}}-\frac{g^2}{\tilde{\delta}} \\
& \ket{UP}=\left( \ket{e,0}-\frac{|g|}{\tilde{\delta}^*}\ket{g,1} \right)N \\
& \ket{LP}=\left( \ket{g,1}+\frac{|g|}{\tilde{\delta}}\ket{e,0} \right)N \\
& N =\frac{1}{\sqrt{1+\frac{g^2}{\delta^2+\Gamma_p^2}}}
\end{split}
\end{equation}

\noindent where we have phenomenologically introduced $\overline{\omega_{p}}=\omega_p-i\Gamma_p$ and so $\tilde{\delta}=\delta+i\Gamma_p$ to explicitly recover the damping rate of the plasmon mode in the FQ model, which also appears in the SC expression of eq.\ref{eq:pop2}. Besides, to ease the notation we set $g_{p}=g$ (which is also real), and given that the molecule-NP distance is large enough to ensure that only dipolar interactions are relevant, the coupling $g$ can be explicitly expressed as
\begin{equation}\label{eq:g}
|g|=\frac{2|\vec{\mu}_p||\vec{\mu}_{e}|}{|\vec{r}\,|^3}
\end{equation}
which is the dipolar coupling between two aligned dipoles. Since the two dipoles point in the same direction, $g<0$, and so the upper state $\ket{UP}$ of eq.\ref{eq:tipt} features the minus combination.
Plugging the results of eq.\ref{eq:tipt} into eq.\ref{eq:ce_fq2} leads to
\begin{equation}\label{eq:ce_fq3}
\begin{split}
C_{e,FQ}^{'}(t)= & \frac{E_0}{2} N \left(|\vec{\mu}_{e}|-\frac{|g|}{\tilde{\delta}}|\vec{\mu}_p|\right)\left( \frac{e^{-i\omega_{p}t}}{\delta+\frac{g^2}{\delta+i\Gamma_p}}+ \frac{e^{i\omega_{p}t}}{2\omega_p+\delta+\frac{g^2}{\delta+i\Gamma_p}}
\right) + \\
& \frac{E_0}{2} N \left(\frac{g^2}{\delta^2+\Gamma_p^2}|\vec{\mu}_{e}|+\frac{|g|}{\tilde{\delta}}|\vec{\mu}_p|\right)\left( \frac{e^{-i\omega_{p}t}}{-i\Gamma_p-\frac{g^2}{\delta+i\Gamma_p}}+ \frac{e^{i\omega_{p}t}}{2\omega_{p}-i\Gamma_p-\frac{g^2}{\delta+i\Gamma_p}}
\right).
\end{split}
\end{equation}

\noindent The expression of eq.\ref{eq:ce_fq3} can be further simplified considering that in the weak-coupling limit we are dealing with, both conditions $g << \delta$ and $g << \Gamma_p$ are satisfied, which implies that $2^{nd}$-order terms like $\frac{g^2}{\delta^2+\Gamma_p^2} \approx 0$ can be safely neglected and $N \approx 1$. It also follows that $\delta+\frac{g^2}{\delta+i\Gamma_p}\approx\delta$ and $-i\Gamma_p-\frac{g^2}{\delta+i\Gamma_p}\approx -i\Gamma_p$ apply too, thus resulting in
\begin{equation}\label{eq:ce_fq4}
C_{e,FQ}^{'}(t)=  \frac{E_0}{2}\left[\left(|\vec{\mu}_{e}|-\frac{g}{\tilde{\delta}}|\vec{\mu}_p|\right)\left( \frac{e^{-i\omega_{p}t}}{\delta}+ \frac{e^{i\omega_{p}t}}{2\omega_p+\delta}
\right)
+\frac{g}{\tilde{\delta}}|\vec{\mu}_p|\left( \frac{e^{-i\omega_{p}t}}{-i\Gamma_p}+ \frac{e^{i\omega_{p}t}}{2\omega_{p}-i\Gamma_p}
\right)\right].
\end{equation}

\noindent Substituting the expressions of eqs.\ref{eq:tipt}-\ref{eq:g} into eq.\ref{eq:ce_fq4} and grouping the corresponding rotating ($\propto e^{-i\omega_pt})$ and counter-rotating field terms ($\propto e^{i\omega_pt})$, we end up with
\begin{equation}\label{eq:ce_fq5}
\begin{split}
C_{e,FQ}^{'}(t)=  \frac{E_0}{2} & \left[ \left(\frac{|\vec{\mu}_{e}|}{\delta}-\frac{2|\vec{\mu}_p|^2|\vec{\mu}_{e}|(\delta-i\Gamma_p)}{|\vec{r}\,|^3 (\delta^2+\Gamma_p^2)\delta} +
\frac{i2|\vec{\mu}_p|^2|\vec{\mu}_{e}|(\delta-i\Gamma_p)}{|\vec{r}\,|^3 (\delta^2+\Gamma_p^2)\Gamma_p}\right)e^{-i\omega_pt}
+\right. \\
&\left. \left(\frac{|\vec{\mu}_{e}|}{2\omega_p+\delta}-\frac{2|\vec{\mu}_p|^2|\vec{\mu}_{e}|}{|\vec{r}\,|^3 \tilde{\delta}(2\omega_p+\delta)}
+\frac{2|\vec{\mu}_p|^2|\vec{\mu}_{e}|}{|\vec{r}\,|^3 \tilde{\delta}(2\omega_p-i\Gamma_p)}\right)e^{i\omega_pt} \right]
\end{split}
\end{equation}

\noindent which after some simple algebraic manipulation leads to
\begin{equation}\label{eq:ce_fq6}
C_{e,FQ}^{'}(t)=  \frac{E_0|\vec{\mu}_{e}|}{2}\left[\left(1+\frac{2i|\vec{\mu}_p|^2}{|\vec{r}\,|^3 \Gamma_p}
\right) \frac{e^{-i\omega_pt}}{\delta}
+\left(1+\frac{2|\vec{\mu}_p|^2}{|\vec{r}\,|^3 (2\omega_p-i\Gamma_p) }\right)\frac{e^{i\omega_pt}}{2\omega_p+\delta}
\right].
\end{equation}

\noindent Taking the squared modulus of eq.\ref{eq:ce_fq6} results in an expression for the molecular excited state population in the FQ picture, that is
\begin{equation}\label{eq:pop_fq}
\begin{split}
|C_{e,FQ}^{'}(t)|^2=  \frac{E_0^2|\vec{\mu}_{e}|^2}{4} & \left[\left(1+\frac{4|\vec{\mu}_p|^4}{|\vec{r}\,|^6 \Gamma_p^2}
\right)\frac{1}{\delta^2}
+\left(1+\frac{4|\vec{\mu}_p|^4}{|\vec{r}\,|^6 (4\omega_p^2+\Gamma_p^2) }+\frac{8|\vec{\mu}_p|^2\omega_p}{|\vec{r}\,|^3 (4\omega_p^2+\Gamma_p^2) }\right) \cdot \right. \\ 
 & \left. \frac{1}{(2\omega_p+\delta)^2}  
+  2\mathfrak{Re}\biggl\{\frac{e^{-i2\omega_{p}t}}{(2\omega_{p}+\delta)\delta} \left(
1+\frac{2i|\vec{\mu}_p|^2}{|\vec{r}\,|^3 \Gamma_p}+\frac{2|\vec{\mu}_p|^2}{|\vec{r}\,|^3 (2\omega_p+i\Gamma_p)}+ \right. \right. \\
& \left. \left. \frac{4i|\vec{\mu}_p|^4}{|\vec{r}\,|^6 \Gamma_p(2\omega_p+i\Gamma_p)}  
\right) \biggl\} \right].
\end{split}
\end{equation}

\noindent The most dominant terms of the SC and FQ expressions (eqs. \ref{eq:pop2} and \ref{eq:pop_fq}, respectively) are those originating from the rotating-wave terms of the incoming field, that is those terms that stem from the field component $\propto e^{-i\omega_pt}$. Therefore,  retaining only those in the corresponding expressions finally brings us to
\begin{equation}\label{eq:pop_comp}
\begin{split}
    & |C_{e,SC}^{'}(t)|^2  \approx \frac{E_{0}^2|\vec{\mu}_{e}|^2}{4} \left( 1 + \frac{4|\vec{\mu}_p|^4}{\Gamma_{p}^2|\vec{r}\,|^6}-\frac{4|\vec{\mu}_p|^4}{(4\omega_{p}^2 +\Gamma_{p}^2)|\vec{r}\,|^6}+
    \frac{8|\vec{\mu}_p|^2\omega_{p}}{(4\omega_{p}^2+\Gamma_{p}^2)|\vec{r}\,|^3} \right)\frac{1}{\delta^2} \\
    & |C_{e,FQ}^{'}(t)|^2 \approx \frac{E_0^2|\vec{\mu}_{e}|^2}{4} \left( 1+\frac{4|\vec{\mu}_p|^4}{\Gamma_p^2|\vec{r}\,|^6 }
\right)\frac{1}{\delta^2}
\end{split}
\end{equation}

\noindent which clearly shows that the small-yet-non-null discrepancy observed in the results of Fig.3 (main text) can be traced back to the anti-resonant term of the polarizability (eq.\ref{eq:alfa3}) which is responsible for the additional terms of $|C_{e,SC}^{'}(t)|^2$ that are absent in  $|C_{e,FQ}^{'}(t)|^2$ in eq.\ref{eq:pop_comp}.

\subsection{Comparison of SC and FQ models under resonance condition}\label{sec:res_pop}

In Fig.4 (main text) it is shown that when the molecular and plasmonic transitions are resonant ($\delta=0$) the FQ and SC molecular excited state population profiles are always perfectly superimposed regardless of the absolute value of $\omega_p$, which is something that can not be easily understood by looking at the expressions of eq.\ref{eq:pop_comp} derived above, since both diverge in this limit. This feature can be actually rationalized considering a more general model still rooted in time-dependent perturbation theory.

In the FQ picture, the full system Hamiltonian in general terms reads
\begin{equation}\label{eq:H_toy}
    \hat{H}_{tot}=\hat{H}+\hat{V}(t)
\end{equation}
\noindent where $\hat{H}$ is the molecule-plasmon Hamiltonian and $\hat{V}(t)=\hat{V}e^{-i\omega t}+\hat{V}^{\dagger}e^{i\omega t}$ is a monochromatic oscillatory perturbation of frequency $\omega$. The shape of $\hat{V}(t)$ guarantees that $\hat{H}_{tot}$ is hermitian\cite{mcweeny}. Starting from eq.\ref{eq:H_toy} and considering the ground state $\ket{g}$ and two molecular and plasmon excited states $\ket{e},\,\ket{p}$, the full system wavefunction at time \textit{t} can be represented as $\ket{\psi(t)}=C_{g}(t)\ket{g}+C_{e}(t)\ket{e}+C_{p}(t)\ket{p}$ and usual time-dependent perturbation theory leads to the following expressions for excited state coefficients, 
\begin{equation}
 \begin{split}
     & i\dot{C}_{e}=\omega_eC_e+H_{ep}C_p+\left(V_{eg}e^{-i\omega t}+V_{ge}^*e^{i\omega t}\right) \\
     & i\dot{C}_{p}=\omega_pC_p+H_{pe}C_e+\left(V_{pg}e^{-i\omega t}+V_{gp}^*e^{i\omega t}\right)
 \end{split}\label{eq:c_dot}
\end{equation}

\noindent where the shorthand notation $\bra{e}\hat{H}\ket{e}=\omega_e$ and $\bra{e}\hat{H}\ket{p}=H_{ep}$ has been introduced for convenience. \\
Given that the monochromatic perturbation has two separate resonant ($\propto e^{-i\omega t}$) and anti-resonant ($\propto e^{i\omega t}$) terms, the molecular and plasmon coefficients can be partitioned accordingly as
\begin{equation}
 \begin{split}
     & {C}_{e}(t)=C_{e+}e^{-i\omega t}+C_{e-}e^{i\omega t} \\
     & {C}_{p}(t)=C_{p+}e^{-i\omega t}+C_{p-}e^{i\omega t}
 \end{split}\label{eq:c_sep}  
\end{equation}
\noindent with $C_{e+}\, (C_{e-})$ representing the resonant (anti-resonant) contribution to the molecular time-dependent excited state coefficient $C_e(t)$. The same goes for $C_p(t)$. \\
Upon differentiation of eqs.\ref{eq:c_sep} and substitution into eqs.\ref{eq:c_dot}, the following relations arise after collecting resonant and anti-resonant terms,
\begin{equation}\label{eq:c_split}
 \begin{split}
     & C_{e+}=-\frac{H_{ep}C_{p+}+V_{eg}}{\omega_e-\omega} \\
     & C_{e-}=-\frac{H_{ep}C_{p-}+V_{ge}^*}{\omega_e+\omega} \\
     & C_{p+}=-\frac{H_{pe}C_{e+}+V_{pg}}{\omega_p-\omega} \\
     & C_{p-}=-\frac{H_{pe}C_{e-}+V_{gp}^*}{\omega_p+\omega} \\
 \end{split}   
\end{equation}
\noindent which interestingly show that resonant and anti-resonant molecular and plasmonic terms do not mix in the FQ model. In other words, the resonant response of the plasmonic system solely determines the resonant response of the molecule and the same goes for the anti-resonant contribution. Eqs.\ref{eq:c_split} can be made more explicit by recognizing that in the investigated case, where the molecular system is very far from the metallic surface, the coupling matrix element $H_{ep}$ can be expressed in terms of plasmonic and molecular transition dipoles (eq.\ref{eq:g}, SI\,\ref{sec:FQ}),  here renamed as $\mu_p^{tr}=\bra{g}\hat{\mu}\ket{p},\, \mu_e^{tr}=\bra{g}\hat{\mu}\ket{e}$,  thus resulting in
\begin{equation}\label{eq:c_split2}
 \begin{split}
     & C_{e+}=-\frac{A_{p+}+V_{eg}}{\omega_e-\omega} \ , \  A_{p+}=\frac{\mu_e^{tr}\mu_p^{tr}C_{p+}}{r^3}\\
     & C_{e-}=-\frac{A_{p-}+V_{ge}^*}{\omega_e+\omega} \ , \  A_{p-}=\frac{\mu_e^{tr}\mu_p^{tr}C_{p-}}{r^3} \\
     & C_{p+}=-\frac{A_{e+}+V_{pg}}{\omega_p-\omega} \ , \  A_{e+}=\frac{\mu_p^{tr}\mu_e^{tr}C_{e+}}{r^3} \\
     & C_{p-}=-\frac{A_{e-}+V_{gp}^*}{\omega_p+\omega} \ , \  A_{e-}=\frac{\mu_p^{tr}\mu_e^{tr}C_{e-}}{r^3} \ . \\
 \end{split}   
\end{equation}

\indent The quantity that is analyzed in Fig.4 (main text), is the molecular excited state population upon driving the system with a field resonant with the plasmon frequency ($\omega=\omega_p$), which can be computed by taking the squared modulus of the molecular coefficient $C_e(t)$ of eq.\ref{eq:c_sep}. Given the expressions of eq.\ref{eq:c_split2} and that excitation is resonant with $\omega=\omega_p$, $C_{e+}$, which depends on the resonant plasmonic term $C_{p+}$, always dominate, even when the molecular and plasmonic transitions are not resonant i.e. $\delta = \omega_e -\omega_p\neq 0$, and so the anti-resonant terms $C_{e-},\, C_{p-}$ are always negligible.

On the other hand, in the SC picture the molecule is still described at quantum-mechanical level but the plasmonic object is classical. In this case the corresponding equations of motion for the molecular and plasmonic systems become\cite{dall2020real} 
\begin{equation}
 \begin{split}
     & i\dot{C}_{e}=\omega_eC_e+\frac{\mu_p\mu_e^{tr}}{r^3}+\left(V_{eg}e^{-i\omega t}+V_{ge}^*e^{i\omega t}\right) \\
     & \ddot{\mu}_p+\mu_p\omega_p^2= K\left(\frac{\mu_e}{r^3}+\left(Ve^{-i\omega t}+V^*e^{i\omega t}\right)\right)
 \end{split}\label{eq:c_dot_sc}
\end{equation}
where $\mu_p$ and $\mu_e$ correspond in the FQ picture to the oscillating first-order plasmonic and molecular contributions to the expectation value of the dipole operator upon driving the coupled system, namely
\begin{equation}
 \begin{split}
     & \mu_p=\left[\mu_p^{tr}C_{p+}+(\mu_p^{tr})^*C_{p-}^*\right]e^{-i \omega t}+\left[\mu_p^{tr}C_{p-}+(\mu_p^{tr})^*C_{p+}^*\right]e^{i \omega t}=\mu_{p+}e^{-i \omega t}+\mu_{p-}e^{i \omega t} \\
     & \mu_e=\left[\mu_e^{tr}C_{e+}+(\mu_e^{tr})^*C_{e-}^*\right]e^{-i \omega t}+\left[\mu_e^{tr}C_{e-}+(\mu_e^{tr})^*C_{e+}^*\right]e^{i \omega t}=\mu_{e+}e^{-i \omega t}+\mu_{e-}e^{i \omega t}.
 \end{split}\label{eq:mu_sc}
\end{equation}
Therefore, $\frac{\mu_p\mu_e^{tr}}{r^3}$ of eq.\ref{eq:c_dot_sc} represents the interaction of the molecular quantum system $\ket{e}$ due to the classical scattered field of the plasmonic dipole $\mu_p$ whose equation of motion is that typical of an harmonic oscillator\cite{dall2020real} driven by the external drive and by the nearby oscillating molecular dipole. The latter term is proportional to $\frac{\mu_e}{r^3}$ and is nothing but the scattered field of dipole $\mu_e$ at the nanoparticle position. K is a numerical factor representing the squared plasma frequency entering into the definition of the metal dielectric function (see ref.\cite{dall2020real} for more details on the classical equation of motion).  \\ Starting from eq.\ref{eq:c_dot_sc} and by applying the same partitioning strategy of eq.\ref{eq:c_sep} to both $C_e(t)$ and $\mu_p$ the following expressions come up
\begin{equation}\label{eq:c_split_sc}
 \begin{split}
     & C_{e+}=-\frac{A_{p+}+V_{eg}}{\omega_e-\omega} \ , \  A_{p+}=\frac{\mu_e^{tr}\mu_{p+}}{r^3}\\
     & C_{e-}=-\frac{A_{p-}+V_{ge}^*}{\omega_e+\omega} \ , \  A_{p-}=\frac{\mu_e^{tr}\mu_{p-}}{r^3}\\
     & \mu_{p+}=(A_{e+}+V)\frac{K}{2\omega_p} \left(\frac{1}{\omega_p-\omega} + \frac{1}{\omega_p+\omega} \right) \ , \  A_{e+}=\frac{\mu_{e+}}{r^3}\\
     & \mu_{p-}=(A_{e-}+V^*)\frac{K}{2\omega_p} \left(\frac{1}{\omega_p-\omega} + \frac{1}{\omega_p+\omega} \right) \ , \  A_{e-}=\frac{\mu_{e-}}{r^3}\\
 \end{split}   
\end{equation}

\noindent with $\mu_{e+}$ and $\mu_{e-}$ defined according to eq.\ref{eq:mu_sc}. \\
Interestingly, by comparing eqs.\ref{eq:c_split2},\ref{eq:c_split_sc} it can be observed that in the FQ model, $C_{e+}$, which represents the most dominant contribution to the molecular excited state population upon driving, solely depends on $C_{p+}$ which is $\propto (\omega_p-\omega)^{-1}$, whereas in the SC picture $C_{e+}$ depends on $\mu_{p+}$ that is $\propto (\omega_p-\omega)^{-1}+(\omega_p+\omega)^{-1}$.   Recalling that external excitation is always resonant with the plasmonic system ($\omega=\omega_p$), it can be observed that under resonance condition ($\delta=0$ and so $\omega=\omega_e=\omega_p$) the most dominant contribution to $C_{e+}$ (and so $C_e$), is numerically equally described by both models since it originates from the component $\propto (\omega_e-\omega)^{-1}(\omega_p-\omega)^{-1}$ that is present in both cases, thus justifying why FQ and SC curves of Fig.4 (main text) under resonance conditions are always superimposed. In other words, in this limit both models treat in the same way the most dominant contribution to the molecular excited state population, which comes from the resonant terms, thus leading to identical results.\\ On the other hand, when the plasmonic and molecular systems are not in resonance ($\delta \neq 0$) also the term $\propto (\omega_p+\omega)^{-1}$ contributing to $C_{e+}$ in the SC model and absent in the corresponding FQ expression plays a small-yet-observable role, as shown in Fig.\,4 (main text). Besides, under this condition also $C_{e-}$ can be significant and this term enters differently in the two cases. More specifically, in the FQ model (eqs.\ref{eq:c_split2}) this contribution only depends on the anti-resonant plasmonic response that is proportional to $\propto \left(\omega_p+\omega\right)^{-1}$, whereas in the SC picture (eqs.\ref{eq:c_split_sc}) $C_{e-}$ depends on $\mu_{p-}$ that in turn has two contributions respectively proportional to $\propto \left(\omega_p-\omega\right)^{-1}$ and $\propto \left(\omega_p+\omega\right)^{-1}$. The first of these two terms, which is absent in the corresponding FQ expression,  is numerically non-negligible because external driving is always resonant with the plasmon frequency, thus constituting an additional plausible source of discrepancy when $\delta \neq 0$.

\section{Scaling NP plasmonic quantities}

In the case of a plasmonic NP described by a Drude-Lorentz (DL) dielectric function model, the NP quantization scheme that has been detailed previously\cite{fregoni2021} leads to the following expressions for plasmon mode frequencies $\omega_p$ and corresponding quantize surface charges $q_{pj}$,
 \begin{equation}\label{eq:DL1}
 \begin{split}
 & \omega_p^2=\omega_0^2+\left(1+\frac{\lambda_p}{2\pi}\right)\frac{\Omega_{plasma}^2}{2}  \\[2ex]
 & q_{pj}\propto\sqrt{\frac{\omega_p^2-\omega_0^2}{2\omega_p}}
\end{split}
\end{equation}

\noindent where $\lambda_p$ is the $p^{th}$ eigenvalue of the NP-PCM diagonalized response function\cite{fregoni2021}, $\omega_0$ is the frequency of DL bound oscillator and $\Omega_{plasma}$ is the metal plasma frequency, as defined in eq.\ref{eq:DL} (SI \ref{sub:semic}). \\
Starting from eq.\ref{eq:DL1} it can be shown that for a given $\lambda_p$ the corresponding plasmon frequency $\omega_p$ can be scaled n-times, with n being a positive integer, while keeping the value of $q_{pj}$ constant if $\omega_0$ and $\Omega_{plasma}$ are respectively multiplied by two real numbers a and b, such that 
 \begin{equation}\label{eq:DL2}
 \begin{split}
 & a^2=\left(\frac{\omega_p^2}{\omega_0^2}\left( n-1 \right)+1\right)n  \\
 & b^2=n\,.
\end{split}
\end{equation}

\noindent Indeed, upon substituting $\omega_0\rightarrow a\omega_0$ and $\Omega_{plasma} \rightarrow \sqrt{n}\,\Omega_{plasma}$ in eq.\ref{eq:DL1} with a,b satisfying the conditions of eq.\ref{eq:DL2} it follows that
 \begin{equation}\label{eq:DL3}
 \begin{split}
 & a^2\omega_0^2+\left(1+\frac{\lambda_p}{2\pi}\right)\frac{b^2\Omega_{plasma}^2}{2} = n^2\omega_p^2 \\[2ex]
 & q_{pj}\propto\sqrt{\frac{\omega_p^2-\omega_0^2}{2\omega_p}}=\sqrt{\frac{n^2\omega_p^2-a^2\omega_0^2}{2n\omega_p}}
\end{split}
\end{equation}

\noindent where the last equality can be easily proved to be true using the relations of eq.\ref{eq:DL2}. \\ 
Eq.\ref{eq:DL3}  shows that with such a,b scaling parameters the plasmon mode frequency is scaled n-times while the corresponding quantized charge does not vary, thus enabling us to perform multiple simulations (Fig.4, main text) where the absolute value of $\omega_p$ is progressively increased while keeping all the other plasmonic quantities of eq.\ref{eq:pop_comp} fixed.

\newpage
\bibliography{ref}